\documentclass[%
 reprint,
 amsmath,
 amssymb,
 aps,
 prd,
 floatfix,
 twocolumn,
 showpacs,
 superscriptaddress
]{revtex4-2}

\usepackage{graphicx}
\usepackage{hyperref}
\usepackage[subrefformat=parens,labelformat=parens,caption=false]{subfig}
\usepackage{multirow}
\usepackage{diagbox}
\usepackage{dblfloatfix}

\begin{document}

\preprint{APS/123-QED}

\title{Measurement of charged-current muon neutrino-argon interactions without
pions in the final state using the MicroBooNE detector}

\newcommand{\ANL}{Argonne National Laboratory (ANL), Lemont, IL, 60439, USA}
\newcommand{\Bern}{Universit{\"a}t Bern, Bern CH-3012, Switzerland}
\newcommand{\BNL}{Brookhaven National Laboratory (BNL), Upton, NY, 11973, USA}
\newcommand{\UCSB}{University of California, Santa Barbara, CA, 93106, USA}
\newcommand{\Cambridge}{University of Cambridge, Cambridge CB3 0HE, United Kingdom}
\newcommand{\CIEMAT}{Centro de Investigaciones Energ\'{e}ticas, Medioambientales y Tecnol\'{o}gicas (CIEMAT), Madrid E-28040, Spain}
\newcommand{\Chicago}{University of Chicago, Chicago, IL, 60637, USA}
\newcommand{\Cincinnati}{University of Cincinnati, Cincinnati, OH, 45221, USA}
\newcommand{\CSU}{Colorado State University, Fort Collins, CO, 80523, USA}
\newcommand{\Columbia}{Columbia University, New York, NY, 10027, USA}
\newcommand{\Edinburgh}{University of Edinburgh, Edinburgh EH9 3FD, United Kingdom}
\newcommand{\FNAL}{Fermi National Accelerator Laboratory (FNAL), Batavia, IL 60510, USA}
\newcommand{\Granada}{Universidad de Granada, Granada E-18071, Spain}
\newcommand{\IIT}{Illinois Institute of Technology (IIT), Chicago, IL 60616, USA}
\newcommand{\ICL}{Imperial College London, London SW7 2AZ, United Kingdom}
\newcommand{\Indiana}{Indiana University, Bloomington, IN 47405, USA}
\newcommand{\Kansas}{The University of Kansas, Lawrence, KS, 66045, USA}
\newcommand{\KSU}{Kansas State University (KSU), Manhattan, KS, 66506, USA}
\newcommand{\Lancaster}{Lancaster University, Lancaster LA1 4YW, United Kingdom}
\newcommand{\LANL}{Los Alamos National Laboratory (LANL), Los Alamos, NM, 87545, USA}
\newcommand{\Louisiana}{Louisiana State University, Baton Rouge, LA, 70803, USA}
\newcommand{\Manchester}{The University of Manchester, Manchester M13 9PL, United Kingdom}
\newcommand{\MIT}{Massachusetts Institute of Technology (MIT), Cambridge, MA, 02139, USA}
\newcommand{\Michigan}{University of Michigan, Ann Arbor, MI, 48109, USA}
\newcommand{\MSU}{Michigan State University, East Lansing, MI 48824, USA}
\newcommand{\Minnesota}{University of Minnesota, Minneapolis, MN, 55455, USA}
\newcommand{\Nankai}{Nankai University, Nankai District, Tianjin 300071, China}
\newcommand{\NMSU}{New Mexico State University (NMSU), Las Cruces, NM, 88003, USA}
\newcommand{\Oxford}{University of Oxford, Oxford OX1 3RH, United Kingdom}
\newcommand{\Pitt}{University of Pittsburgh, Pittsburgh, PA, 15260, USA}
\newcommand{\QMUL}{Queen Mary University of London, London E1 4NS, United Kingdom}
\newcommand{\Rutgers}{Rutgers University, Piscataway, NJ, 08854, USA}
\newcommand{\SLAC}{SLAC National Accelerator Laboratory, Menlo Park, CA, 94025, USA}
\newcommand{\SDSMT}{South Dakota School of Mines and Technology (SDSMT), Rapid City, SD, 57701, USA}
\newcommand{\Maine}{University of Southern Maine, Portland, ME, 04104, USA}
\newcommand{\TelAviv}{Tel Aviv University, Tel Aviv, Israel, 69978}
\newcommand{\UTA}{University of Texas, Arlington, TX, 76019, USA}
\newcommand{\Tufts}{Tufts University, Medford, MA, 02155, USA}
\newcommand{\VTech}{Center for Neutrino Physics, Virginia Tech, Blacksburg, VA, 24061, USA}
\newcommand{\Warwick}{University of Warwick, Coventry CV4 7AL, United Kingdom}

\affiliation{\ANL}
\affiliation{\Bern}
\affiliation{\BNL}
\affiliation{\UCSB}
\affiliation{\Cambridge}
\affiliation{\CIEMAT}
\affiliation{\Chicago}
\affiliation{\Cincinnati}
\affiliation{\CSU}
\affiliation{\Columbia}
\affiliation{\Edinburgh}
\affiliation{\FNAL}
\affiliation{\Granada}
\affiliation{\IIT}
\affiliation{\ICL}
\affiliation{\Indiana}
\affiliation{\Kansas}
\affiliation{\KSU}
\affiliation{\Lancaster}
\affiliation{\LANL}
\affiliation{\Louisiana}
\affiliation{\Manchester}
\affiliation{\MIT}
\affiliation{\Michigan}
\affiliation{\MSU}
\affiliation{\Minnesota}
\affiliation{\Nankai}
\affiliation{\NMSU}
\affiliation{\Oxford}
\affiliation{\Pitt}
\affiliation{\QMUL}
\affiliation{\Rutgers}
\affiliation{\SLAC}
\affiliation{\SDSMT}
\affiliation{\Maine}
\affiliation{\TelAviv}
\affiliation{\UTA}
\affiliation{\Tufts}
\affiliation{\VTech}
\affiliation{\Warwick}

\author{P.~Abratenko} \affiliation{\Tufts}
\author{D.~Andrade~Aldana} \affiliation{\IIT}
\author{L.~Arellano} \affiliation{\Manchester}
\author{J.~Asaadi} \affiliation{\UTA}
\author{A.~Ashkenazi}\affiliation{\TelAviv}
\author{S.~Balasubramanian}\affiliation{\FNAL}
\author{B.~Baller} \affiliation{\FNAL}
\author{A.~Barnard} \affiliation{\Oxford}
\author{G.~Barr} \affiliation{\Oxford}
\author{D.~Barrow} \affiliation{\Oxford}
% ...

\collaboration{The MicroBooNE Collaboration}
\thanks{microboone\_info@fnal.gov}\noaffiliation

\begin{abstract}
We report a new measurement of flux-integrated differential cross sections for
charged-current (CC) muon neutrino interactions with argon nuclei that produce
no final-state pions ($\nu_\mu\mathrm{CC}0\pi$). These interactions are of
particular importance as a topologically defined signal dominated by
quasielasticlike interactions. This measurement was performed with the
MicroBooNE liquid argon time projection chamber detector located at the Fermilab
Booster Neutrino Beam and uses an exposure of $1.3\times10^{21}$ protons
on target collected between 2015 and 2020. The results are presented in terms of
single- and double-differential cross sections as a function of the final-state
muon momentum and angle. The data are compared with widely used neutrino event
generators. We find good agreement with the single-differential measurements,
while only a subset of generators are also able to adequately describe the data
in double-differential distributions. This work facilitates comparison with
Cherenkov detector measurements, including those located at the Booster Neutrino Beam.
\end{abstract}

\maketitle

\section{Introduction}
\label{sec:intro}
Accelerator-based neutrino experiments enable precise measurements of
oscillation parameters, aiming to conclusively determine the unknown neutrino
mass ordering and establish the presence of charge-parity symmetry violation in
the lepton sector~\cite{Branco:2011zb, DeSalas:2018rby}. These future
measurements require unprecedented precision in the characterization of
neutrino-nucleus interactions for a variety of target nuclei and topological
event signatures. Liquid argon time projection chamber (TPC)-based experiments such as the Fermilab
Short-Baseline Neutrino Program~\cite{MicroBooNE:2015bmn, Machado:2019oxb}
and DUNE~\cite{DUNE:2020ypp, DUNE:2020jqi, DUNE:2020fgq, DUNE:2020zfm} take advantage of
low-threshold proton reconstruction to define a charged-current quasielastic
(CCQE)-like signal with one lepton and one proton in the final state
($1\ell1p$). Water Cherenkov experiments such as the Super-Kamiokande far
detector for T2K \cite{T2K:2011qtm, T2K:2019bcf} and Hyper-Kamiokande
\cite{Hyper-Kamiokande:2018ofw, Hyper-KamiokandeProto-:2015xww} typically probe
this channel using a single-ring CC$0\pi$ topology. Such measurements are not
sensitive to the proton multiplicity due to the high Cherenkov threshold for
protons. Hybrid scintillator detectors offer Cherenkov ring--based event
reconstruction but with the addition of calorimetry for subthreshold
hadrons~\cite{Theia:2019non}.

The present work aims to inform modeling and facilitate comparison across
nuclear targets by selecting a high-statistics topological CC$0\pi$
semi-inclusive event sample within the MicroBooNE liquid argon time projection
chamber (LArTPC), including events with no final-state protons to probe the
lepton kinematics associated with these interactions. This work also enables
correlated multitarget cross-section measurements using Cherenkov detectors in
the same neutrino beam, including ANNIE~\cite{ANNIE:2015inw, ANNIE:2017nng}.
This measurement builds on earlier MicroBooNE measurements of CC$0\pi Np$
interactions, which require at least one reconstructed proton in the final
state. The previous measurements present detailed studies of the scattering
kinematics with respect to the muon trajectory~\cite{MicroBooNE:2024tmp,
MicroBooNE:2023krv, MicroBooNE:2023cmw}. Relative to previous work, this
measurement relaxes the final-state proton requirement to expand the sample to
include $\nu_\mu$CC$0\pi$ interactions with no associated protons. The muon
momentum range is also expanded to include muons up to 2\,GeV/$c$. Additionally,
this work uses the full MicroBooNE dataset with an exposure of
$1.3\times10^{21}$ protons on target (POT), approximately double that of
previous $\nu_\mu$CC$0\pi Np$ measurements~\cite{MicroBooNE:2020akw,
MicroBooNE:2023tzj, MicroBooNE:2023cmw, MicroBooNE:2023krv, MicroBooNE:2024tmp}.

The results are presented in terms of single- and double-differential
flux-integrated cross sections as a function of the final-state muon momentum
($p_\mu$) and the cosine of the scattering angle with respect to the incoming
neutrino beam ($\cos\theta_\mu$). The absolute cross sections per argon nucleus
are unfolded from reconstructed distributions into true regularized
distributions using the Wiener-SVD (singular value decomposition) unfolding technique \cite{Tang:2017rob},
which enables hypothesis testing without loss of statistical power relative to
the distributions of reconstructed quantities.

\section{The MicroBooNE experiment}
\label{sec:experiment}
MicroBooNE is a LArTPC detector located at Fermi National Accelerator
Laboratory (Fermilab) that collected data from 2015 to 2020
\cite{MicroBooNE:2016pwy}. MicroBooNE contained approximately 85 metric tons of active
liquid argon within a TPC with dimensions of 2.56\,m horizontally (drift
direction), 10.36\,m along the beam direction, and 2.32\,m vertically. A uniform
273\,V/cm drift field was applied across the TPC active volume. The anode
readout assembly features three sense wire planes with a 3\,mm pitch, used to
detect ionization charge. The sense wire waveforms provide measurements of
energy deposition and particle trajectories, enabling 3D track reconstruction,
momentum estimation, and particle identification. Located behind the anode plane
are 32 photomultiplier tubes covered by acrylic disks coated with a
wavelength shifter. The photomultiplier tubes capture scintillation light from the interactions,
providing nanosecond-level timing that associates reconstructed tracks with the
neutrino beam spill. This timing information facilitates matching of TPC charge
and prompt scintillation light signals, allows cosmic-ray activity to be
identified and removed, and provides the location along the drift direction
(transverse to the beam) for beam-related activity.

\subsection{Booster neutrino beam}
\label{sec:microboone:bnb}
The MicroBooNE detector is positioned on-axis 463\,m downstream of the Fermilab
Booster Neutrino Beam (BNB) target and approximately $8^\circ$ off-axis relative
to the Neutrinos at the Main Injector beam line. Only BNB events are
considered in this work. To produce the neutrino beam, protons accelerated to
8\,GeV impinge on a beryllium target resulting in the production of hadrons,
mainly pions. For the entirety of the MicroBooNE run period, the BNB
hadron-focusing magnetic horn system operated in a mode where $\pi^+$ were
focused. These hadrons decay in flight in a 50-m-long decay pipe, leading to a
neutrino flux of predominantly $\nu_\mu$ (93.7$\%$) and a smaller
$\bar{\nu}_\mu$ contribution (5.8$\%$), with the remainder of the flux
comprising $\nu_e$ and $\bar\nu_e$. The average energy of muon neutrinos is
approximately 0.8\,GeV. The simulation and modeling of BNB flux and assessment
of the associated systematic uncertainty follow the procedures previously
developed by MiniBooNE~\cite{MiniBooNE:2008hfu}.

\subsection{Neutrino interaction modeling}
\label{sec:microboone:basemodel}
To model and simulate neutrino-nucleus interactions, we use GENIE version
3.0.6~\cite{GENIE:2021npt,Andreopoulos:2009rq,Andreopoulos:2015wxa} using the
comprehensive model configuration (CMC) {G18\_10a\_02\_11a}
\cite{GENIE:2021zuu}. In this configuration, the nuclear ground state is modeled
using a local Fermi gas approach~\cite{Nieves:2011pp}. CCQE interactions
producing a muon and a single proton ($1p1h$) are treated using the Valencia
model~\cite{Nieves:2004wx, Nieves:2011pp, Nieves:2012yz}. This model also
incorporates multinucleon interactions, dominated by two correlated nucleons,
known as two-particle--two-hole ($2p2h$) interactions, based on the Nieves-Simo-Vacas
$2p2h$ framework~\cite{Nieves:2011pp, Gran:2013kda}. Additionally, the
model accounts for long-range correlations in the nuclear medium using the
random phase approximation~\cite{Nieves:2012yz}. \mbox{GENIE} simulates
single-pion production via heavy baryon decay, known as resonance production, using the Feynman-Kislinger-Ravndal formalism
\cite{Copley:1971cey}, a relativistic three-quark bound-state model. This
approach uses the Berger and Sehgal~\cite{Berger:2007rq, Graczyk:2007bc}
and Kuzmin-Lyubushkin-Naumov~\cite{Nowak:2009se, Kuzmin:2003ji} models,
which incorporate contributions from Rein and Sehgal~\cite{Rein:1980wg}.
Single-pion production involving the entire nucleus as a single entity, with no
nucleon-level breakup or residual excitation, is defined as coherent pion
production and is modeled using the formalism developed by Berger and
Sehgal~\cite{Berger:2008xs}. Deep inelastic scattering, which describes
interactions with individual quarks, is simulated using the Bodek-Yang model~\cite{Bodek:2003wc, Bodek:2004pc, Bodek:2009jmt}. The Bodek-Yang model is a
phenomenological approach that incorporates tuned parton distribution functions extracted from global fits at high $Q^{2}$. It extrapolates these parton distribution functions to
low $Q^{2}$ while accounting for target mass corrections, nonperturbative QCD
effects, and higher-order QCD terms. Final-state interactions,
which occur as hadrons exit the nucleus, are simulated using the hA2018
model~\cite{Dytman:2021ohr}. Importantly, these final-state interactions can lead to CC$0\pi$ final-state topologies from several processes, including, for example, resonance
production interactions where a charged pion is produced and absorbed in the
nucleus.

The {G18\_10a\_02\_11a} CMC is tuned to neutrino interaction data as described
in Ref.~\cite{GENIE:2021zuu}. MicroBooNE's neutrino interaction model uses this
tuned CMC as a starting point and then applies an additional model parameter tuning
based on fits to T2K CC$0\pi$ data~\cite{T2K:2016jor}, used to obtain the
central value and updated uncertainties~\cite{MicroBooNE:2021ccs}. This
``MicroBooNE tune'' of the CC model parameters to an external dataset corrects
for an underprediction of the total CC cross section for $E_\nu\sim1$\,GeV
observed relative to both T2K and MicroBooNE data.

\subsection{Event simulation}
The propagation of simulated final-state particles through the MicroBooNE
detector geometry is performed using Geant4 version 10.3.3~\cite{GEANT4:2002zbu,
Allison:2016lfl}. The detector response is modeled with a custom simulation
implemented within the LArSoft framework~\cite{Snider:2017wjd}. The simulation
and data processing include detailed models of signal processing
\cite{MicroBooNE:2018swd} and wire waveform induction effects
\cite{MicroBooNE:2018vro}, as well as data-driven calibrations and models for
components such as electric field maps~\cite{MicroBooNE:2019koz} and space
charge effects~\cite{MicroBooNE:2020kca}. Monte Carlo neutrino interactions are
overlaid with cosmic-ray data events taken from nonbeam data to account for the
cosmic-ray background in an unbiased way and provide a data-driven model of
detector noise~\cite{MicroBooNE:2018vxr}. Additionally, backgrounds due to
cosmic-ray activity occurring in time with the beam, but where no neutrino
interaction occurred, are estimated using data samples collected with the beam
off (``EXT BNB'').

\subsection{Event reconstruction}
\label{sec:eventreconstuction}
As with previous MicroBooNE CC$0\pi Np$ measurements~\cite{MicroBooNE:2020akw,
MicroBooNE:2023tzj, MicroBooNE:2023cmw, MicroBooNE:2023krv, MicroBooNE:2024tmp},
this work employs algorithms and tools implemented within the Pandora
multialgorithm framework \cite{MicroBooNE:2017xvs} to reconstruct cosmic-ray
muon and neutrino events in the MicroBooNE detector. The algorithms include
three-dimensional track reconstruction, which allows the determination of
kinematic variables such as muon momentum and angle. Pandora begins by grouping
charge hits on individual 2D readout planes and then matches these hits across
planes to form an image of the ionization charge distribution in 3D. Clustering
is applied to form tracks, which are used to identify and remove cosmic-ray
activity. For the remaining neutrino interaction candidates, Pandora identifies
and reconstructs particles produced in the neutrino interaction. These are
characterized as tracklike (e.g., muons, protons, and charged pions) or
electromagnetic-shower-like (e.g., electrons and photon-induced showers). These
tracks and showers are then grouped into a particle flow hierarchy associated
with a reconstructed neutrino vertex.

The energy and momentum of each final-state particle object is computed using
several approaches. For tracklike particles, the momentum
is estimated using the visible track length under the hypotheses that its energy
loss profile is that of a muon or proton. A second momentum estimator uses
multiple Coulomb scattering (MCS), assuming a muon track. This algorithm makes
use of the fact that lower-momentum tracks tend to undergo more MCS, leading to
a higher local curvature~\cite{MicroBooNE:2017tkp}. In this work, we consider
only muon tracks that are fully contained within the detector (i.e., the full
extent of the track is visible in the TPC) and use the total track length
estimator for the momentum. For showers, energy is estimated using the total
deposited charge, with corrections applied for known charge losses due to hit
thresholding and clustering efficiencies.

In addition to the daughter track start and end points, lengths, and momenta, we
also compute a variety of calorimetric quantities using the deposited charge
distribution, which are used for particle identification. Specifically, the
energy loss profile ($dE/dx$) as the track approaches its end point provides a
powerful means to discriminate muon and pion tracks from highly ionizing proton
tracks. Two specific metrics used in this work are a $\chi^2$ score for a track
under the hypothesis that it is a proton ($\chi^2_p$) and a ratio of
likelihoods for a track's $dE/dx$ profile under a muon or proton hypothesis~\cite{MicroBooNE:2021ddy}, which are among the inputs to the particle
identification boosted decision tree (BDT) model described in
Sec.~\ref{sec:BDT}.

\section{Event selection}
\label{sec:selection}

\subsection{Signal and background definitions}
\label{sec:signaldefinition}
This measurement uses a topological CC$0\pi$ signal definition, which refers to
any $\nu_\mu$ CC scattering events with no charged or neutral pions in the final
state. In practice, these events can be identified only when the muon and
charged pion have sufficient kinetic energy to be reliably reconstructed. Thus,
the signal definition includes a muon and charged pion momentum threshold that
is related to the reconstruction threshold. An upper threshold (2\,GeV/$c$) is
also applied to the muon momentum due to the requirement that muons be fully
contained within the detector. This requirement reduces the efficiency
significantly for high-momentum muon tracks that tend to exit the detector.

The signal is defined to include events where (i) a muon neutrino underwent a
CC interaction with an $^{40}\textrm{Ar}$ nucleus, (ii) the final-state muon momentum is in the range $0.1<p_\mu<2.0$\,GeV/$c$, (iii) there are no
neutral pions in the final state, and (iv) there are no charged pions with
momentum $p_\pi>70$\,MeV/$c$.

Simulated events that pass this signal definition and have a vertex within the
fiducial volume (FV) are considered signal events. The FV is an inner
rectangular volume within MicroBooNE's active volume, inset 21.5\,cm from the
TPC boundaries with the exception of the downstream end along the beam axis,
where it is inset 70\,cm from the downstream TPC boundary. For the purposes of
presentation in the figures in later sections, signal events are further labeled
by their true interaction types according to the GENIE generator: CC
quasielastic (CCQE), CC meson exchange currents or $2p2h$ (CCMEC), CC resonance
production (CCRES), and all other CC interaction processes that fall within the
signal definition [``Signal (Other)''], a category dominated by deep inelastic scattering and CC
coherent pion events.

All other events are considered background and grouped into categories: out of
fiducial volume events (``Out FV'') with a true neutrino-induced vertex position
outside of the fiducial volume; neutral current events (``NC'') where, for
example, a proton or pion is misidentified as a muon; electron neutrino CC
(``$\nu_{e}$ CC'') events; other $\nu_{\mu}$ CC interactions (``other $\nu_{\mu}$
CC'') induced by a muon neutrino and having no final-state pions but still
failing to meet the signal definition, for example, with a true muon momentum outside
the signal range; CC events with pions above threshold (``CC$N\pi$''); and
cosmic-ray activity (``EXT BNB'') misidentified as a neutrino interaction. A final
category (``other'') includes any remaining backgrounds including, for example,
antineutrino interactions.

The selection proceeds in three stages. First, the $\nu_{\mu}$CC preselection
provides a sample of CC interaction candidates with reduced contamination from
cosmic-ray activity and electromagnetic showers. Next, this sample is input to a
particle identification algorithm that estimates the final-state track content
of each interaction. Finally, events meeting the topological criteria are
subject to kinematic selections matching those in the signal definition to yield
the final event sample.

\subsection{Muon neutrino CC preselection}
\label{sec:preselection}
We apply a semi-inclusive preselection to the simulation and data to identify
$\nu_{\mu}$ CC events, reducing backgrounds related to cosmic-ray and
electromagnetic shower activity prior to a more targeted signal selection. We
require that Pandora has reconstructed a single neutrino vertex within the FV.
To ensure high-quality reconstruction of particle types and momenta, all tracks
resulting directly from that interaction must start within a particle
containment volume (PCV) inset 10\,cm from the TPC active volume boundary. To
minimize cosmic-ray activity, the Pandora reconstruction provides a likelihood
score that discriminates cosmiclike from neutrinolike event topologies. It is
used to reject topologically cosmiclike events. Another discrimination
algorithm provides a score (0--1) to differentiate between showerlike (closer
to 0) or tracklike (closer to 1) reconstructed final-state objects based on the
spatial distribution of associated clustered hits. A threshold of 0.5 is used to
reject any event containing showerlike activity produced directly from the
neutrino interaction. This suppresses many classes of background events, such as
those with final-state $\pi^0$ mesons.

In contrast to the CC$0\pi Np$ topology where a muon and at least one proton
form a well-identified vertex, the more inclusive CC$0\pi$ channel includes
interactions where the only reconstructed final-state track is that of the muon.
Using topological information alone (i.e., the distribution of ionization hits
in 3D space) creates a $180^\circ$ ambiguity in the track orientation, leading
to an increased acceptance of cosmic-ray muon tracks that stop in the detector
volume. To reduce this background contribution, additional calorimetric
information is used in the muon candidate track selection. Specifically, tracks
are excluded when there is a preference for a Bragg peak at the track end point
closer to the reconstructed neutrino vertex and a poor goodness of fit for the
$dE/dx$ hypothesis corresponding to a forward-going muon.

\subsection{Track identification BDT}
\label{sec:BDT}
The identification of the CC$0\pi$ signal topology is based on the
classification of tracks associated with a reconstructed neutrino interaction.
This classification is performed using an XGBoost~\cite{Chen:2016btl} gradient
boosted decision tree model trained on reconstructed simulated events that pass
the preselection. The input parameters for the model include a set of
reconstructed object parameters computed by the Pandora algorithms
\cite{MicroBooNE:2017xvs}. The objective of the model is to correctly classify
final-state particle objects as muons, charged pions, protons, or none of these
types. The input parameters build on existing discriminants for protons and
muons to construct an optimal classifier. Parameters must fall within a defined
range, establishing loose cuts on the tracks to which the BDT is applied. The
BDT input parameters and their ranges are: distance from the neutrino vertex to
the track start point ($\leq100$\,cm); a muon and proton likelihood ratio and
$\chi^2$ score for the track $dE/dx$ evolution across the track length;
range-based reconstructed track kinetic energy assuming a proton track
($\leq2.5$\,GeV); the fractional difference between the muon momentum as
reconstructed using track range, $p_\mu^\mathrm{range}$,
and an estimator $p_\mu^\mathrm{MCS}$ based on multiple Coulomb
scattering
[$\left|(p_\mu^\mathrm{MCS}-p_\mu^\mathrm{range})/p_\mu^\mathrm{MCS}\right|<2.5$];
track end point containment within the PCV; and
multiplicity of associated tracklike and showerlike secondary particles
($\leq10$).

To validate the BDT model performance, statistical comparisons are performed to
assess the agreement of simulation to a subset of the data corresponding to an
exposure of $1.4\times10^{20}$\,POT. These comparisons include the track-level
distributions of all model input parameters listed above, as well as the
distributions of probability scores for each output class, for all tracklike
reconstructed objects associated with preselected events. Data and simulation
are consistent in all cases, with agreement well within the $1\sigma$ model
uncertainties (enumerated in Sec.~\ref{sec:uncert}).

For events passing the $\nu_\mu$CC preselection, the particle identification
(PID) BDT model provides a probability score for each output class for each
final-state track in a neutrino interaction event. If any final-state track's
parameters fall outside the range of validity for the BDT model inputs, the
event is rejected. The positive identification of a muon candidate track is an
input to the $\nu_\mu$CC selection. Muons are correctly identified in 82\% of
cases. The primary source of muon misidentification is charged pions, which
leave similar signatures in the LArTPC. According to simulations of particles
identified as pions, 56\% are true pions while 22\% are true muons and 13\% are
protons. This nontrivial pion confusion is partly mitigated by the topological
selection. Since each $\nu_\mu$CC candidate event must contain exactly one muon,
the muonlike track with the highest muon BDT score is chosen as the primary
muon candidate. Any remaining tracks that were initially classified as muons are
reclassified as charged pions.

\subsection{Signal selection criteria}
\label{sec:selectionCriteria}
The final selection criteria for the CC$0\pi$ sample are designed to coincide
with the signal definition stated in Sec.~\ref{sec:signaldefinition}, using
reconstructed observables. Overall, we require that (i) events pass the
preselection criteria defined in Sec.~\ref{sec:preselection}; (ii) the BDT
PID model has identified a muon candidate track that is fully contained within
the PCV and has $0.1<p_\mu^\mathrm{reco}<2.0$\,GeV/$c$; (iii) no charged pion tracks
with momentum $p_\pi^\mathrm{reco}>70$\,MeV/$c$ have been identified by the BDT
PID model; (iv) all final-state particles have been identified and consist of
one muon and any number of proton candidate tracks that all start and end within
the PCV; (v) all final-state activity has been classified by the BDT PID model.
This selection requires that all final-state particles be positively identified
by the BDT PID model and pass momentum requirements, improving the purity of the
sample.

\begin{figure}
\subfloat[Efficiency.]{%
\includegraphics[width=0.48\textwidth]{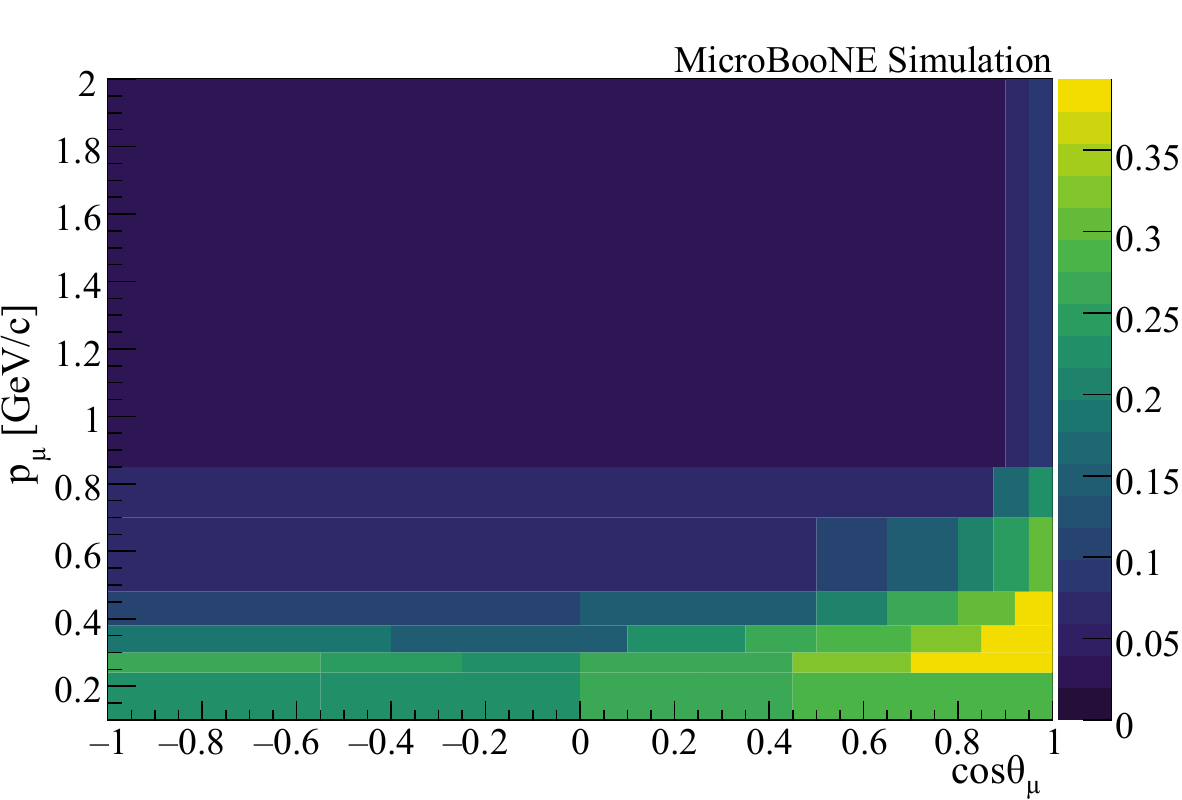}}\hfill
\subfloat[Purity.]{%
\includegraphics[width=0.48\textwidth]{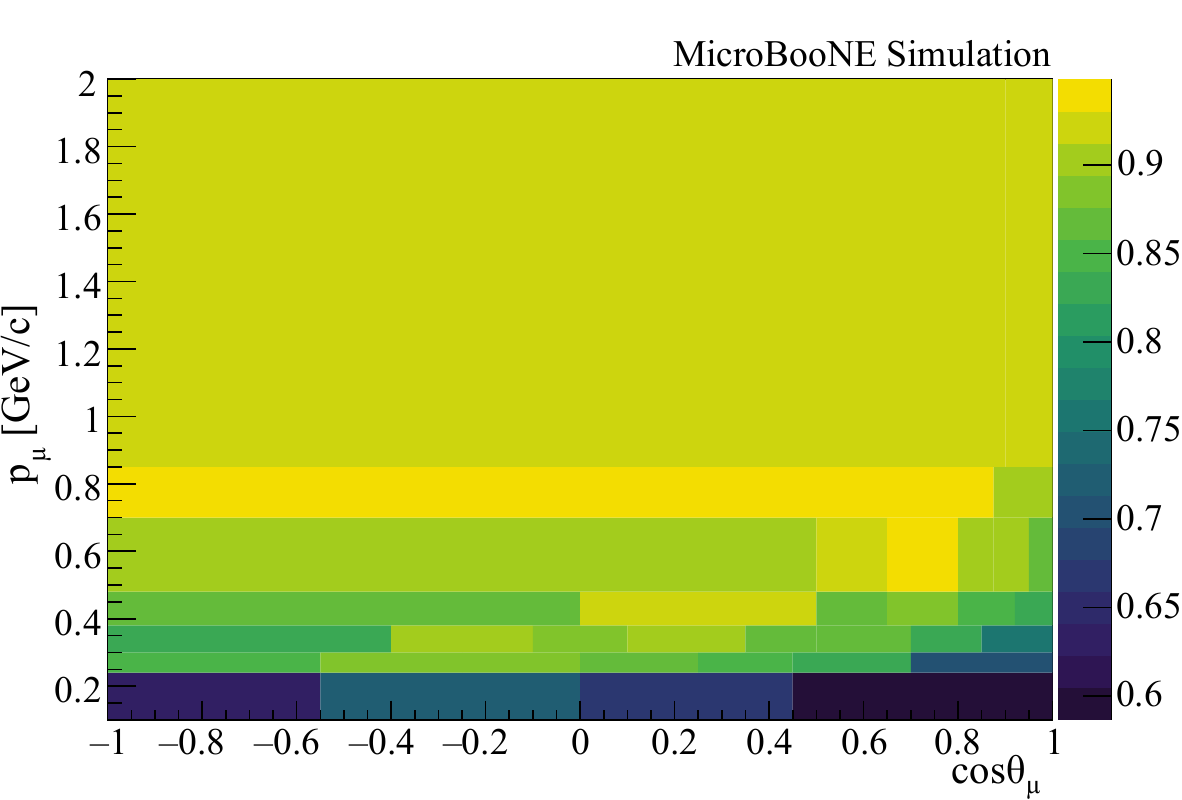}}
\caption{Performance of the $\nu_\mu$CC$0\pi$ event selection, as a function of
muon $p_\mu$ and $\cos\theta_\mu$, using the 2D binning.}
\label{fig:2d-efficiency}
\end{figure}

The CC$0\pi$ selection achieves an overall efficiency of $13\%$ with a sample
purity of $71\%$ (Table~\ref{table:selectionPerformance}). The final requirement
listed is containment of the full extent of the candidate muon within the PCV.
This enables an accurate estimate of the muon momentum using the track's
well-measured range and the known energy loss profile. It is also possible to
measure the momentum of exiting muons using the deflection angles due to
multiple Coulomb scatterings along the track \cite{MicroBooNE:2017tkp}. This
extension to higher muon momentum is not considered in this work. As seen when
comparing the last two rows in Table~\ref{table:selectionPerformance},
containment has a sizable impact on the selection efficiency (37\% $\to$ 13\%)
which particularly impacts higher-momentum muons. Figure~\ref{fig:2d-efficiency}
illustrates the efficiency and purity as a function of muon momentum and angle,
indicating the uniformly high purity. The efficiency is maximal for
intermediate-momentum forward-going muons.

\begin{table}
\caption{Efficiency and purity for each stage in the CC$0\pi$ selection.
The efficiency is shown as an absolute fraction and relative to the
previous step (in parentheses).}
\label{table:selectionPerformance}
\centering
\begin{tabular}{|c|c|c|}
\hline
{Selection criterion} & {Efficiency (rel.)} & {Purity} \\ 
\hline
\hline
            One reconstructed $\nu$ &  0.85 (85\%) &  0.11 \\
\hline
                     Vertex in FV &  0.78 (93\%) &  0.25 \\
\hline
           Neutrinolike topology &  0.65 (82\%) &  0.39 \\
\hline
                       No showers &  0.53 (82\%) &  0.51 \\
\hline
           Daughters start in PCV &  0.52 (99\%) &  0.52 \\
\hline
$\nu_\mu$CC (BDT $\mu$ candidate) &  0.49 (95\%) &  0.57 \\
\hline
                   $p_\mu$ limits &  0.49 (98\%) &  0.58 \\
\hline
              $\pi^\pm$ threshold &  0.44 (90\%) &  0.71 \\
\hline
                 Complete BDT PID &  0.37 (85\%) &  0.74 \\
\hline
                  $\mu$ contained &  0.13 (34\%) &  0.71 \\
\hline
\end{tabular}
\end{table}

\subsection{Binning of observables}
\label{sec:binning}
The flux-integrated cross section is extracted as a function of final-state muon
momentum ($p_\mu$) and the cosine of the muon angle with respect to the beam
axis ($\cos\theta_\mu$). For each observable, the binning choice has been
optimized to ensure adequate event statistics. Each bin in the space of
reconstructed observables is required to contain at least 50 selected signal
events. To accurately reflect the measurement resolution, each reconstructed bin
is required to contain $\geq50\%$ of the events in the corresponding true bin.
The diagonal elements of a normalized smearing matrix are therefore required to
exceed 0.5. The total uncertainty within each reconstructed bin is less than
30\% and is approximately Gaussian. The optimal bin boundaries have been
rounded to the nearest 5~MeV$/c$ and $\cos\theta=0.005$.

Results are presented in two formats: (a) two uncorrelated 1D measurements in
$p_\mu$ and $\cos\theta_\mu$ and (b) a single 2D measurement binned in the
joint space of both observables. In the latter case, the bins are defined to
enable a projection into 1D $p_\mu$ bins. In both cases, bin boundaries cover
the range $0.1<p_\mu<2.0$\,GeV/$c$ and $-1<\cos\theta_\mu<1$. The migration
matrices and numerical values for all bin boundaries are given in
Supplemental Material~\cite{supplement}.

\nocite{Golan:2012rfa}
\nocite{Hayato:2021heg}
\nocite{Bradford:2006yz}
\nocite{Bodek:2007ym}
\nocite{Benhar:1994hw}
\nocite{Golan:2012wx}
\nocite{Leitner:2006ww}
\nocite{Blaettel:1993uz}
\nocite{Sjostrand:2006za}
\nocite{Mosel:2023zek}
\nocite{Mosel:2019vhx}
\nocite{Meyer:2016oeg}
\nocite{Schwehr:2016pvn}
\nocite{Gonzalez-Rosa:2022ltp}
\nocite{LlewellynSmith:1971uhs}
\section{Systematic uncertainties}
\label{sec:uncert}
We consider systematic uncertainties from the simulation and modeling used to
estimate efficiencies and backgrounds. The statistical uncertainties due to the
finite sample size of data-driven background estimates are also treated as
systematic uncertainties. This work follows the same approach as previous
MicroBooNE neutrino cross-section measurements~\cite{Gardiner:2024gdy}. To
provide a sense of relative scale, we note throughout this discussion the
approximate fractional uncertainty associated with each source of uncertainty
near $p_\mu=0.5$\,GeV/$c$. The total uncertainty in this 1D momentum bin is
approximately 10\%, including a 3\% statistical contribution. The fractional
uncertainty in each unfolded cross-section bin can be found in Supplemental
Material~\cite{supplement}.

Uncertainties related to the BNB neutrino flux arise from underlying uncertainty
in the beam simulation, details of the magnetic focusing horn, and hadron
interaction cross sections. They impact the expected rate of neutrinos per POT
and their energy distribution. These uncertainties are propagated by sampling
1000 randomly drawn parameter sets from the correlated distributions of the
underlying simulation parameters to reweight events, following the approach
developed by MiniBooNE~\cite{MiniBooNE:2008hfu}. Flux uncertainties vary within
(5--20)\% across the momentum range, with a minimum of 5\% near $0.5$\,GeV/$c$.
An additional small uncertainty of 2\% accounts for limitations of the POT
estimation. 

The GENIE neutrino interaction model, as constrained in the MicroBooNE tune,
includes 44 parameters. Similar to the flux, 500 randomly drawn variations of
these parameters constitute alternative ``universes'' used to estimate
uncertainties and correlations by reweighting the nominal Monte Carlo
events~\cite{MicroBooNE:2021ccs,MicroBooNE:2023cmw}. Additional variations
account for uncertainties on the vector and axial form factors used by the CCQE
interaction models. The total neutrino interaction uncertainty varies from
(4--23)\% across the momentum range. At $0.5$\,GeV/$c$ it is approximately 5\%,
comparable to flux uncertainties.

Uncertainties related to final-state hadron reinteractions in the bulk argon
(after leaving the nucleus) are implemented using the Geant4Reweight
\cite{Calcutt:2021zck} package. This uncertainty is approximately 1\% throughout
the measured momentum range.

\begin{figure}[htbp]
\subfloat[Reconstructed $p_\mu$.
\label{fig:1d-event-rates:pmu}]{%
\includegraphics[width=0.48\textwidth]{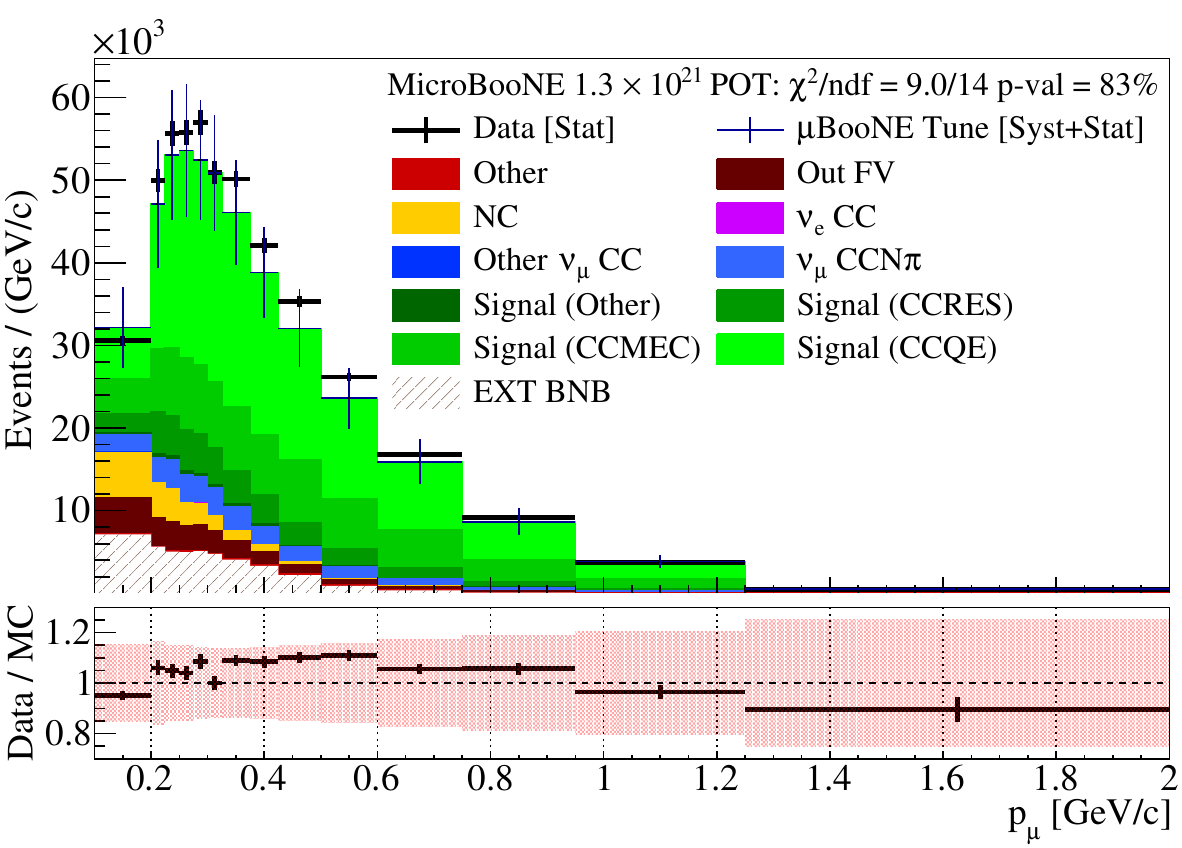}}\\
\subfloat[Reconstructed $\cos\theta_\mu$.
\label{fig:1d-event-rates:ctmu}]{%
\includegraphics[width=0.48\textwidth]{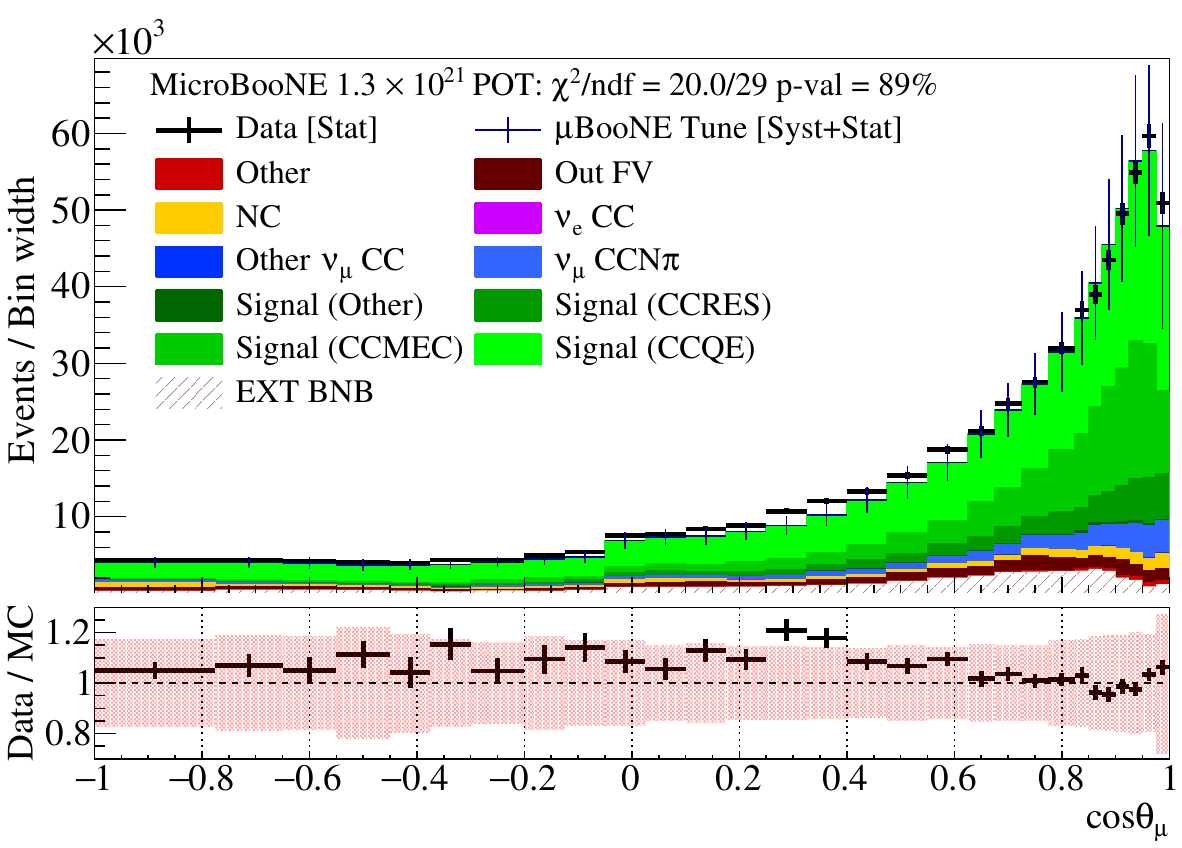}}\\
\subfloat[Reconstructed proton multiplicity.
\label{fig:1d-event-rates:np}]{%
\includegraphics[width=0.48\textwidth]{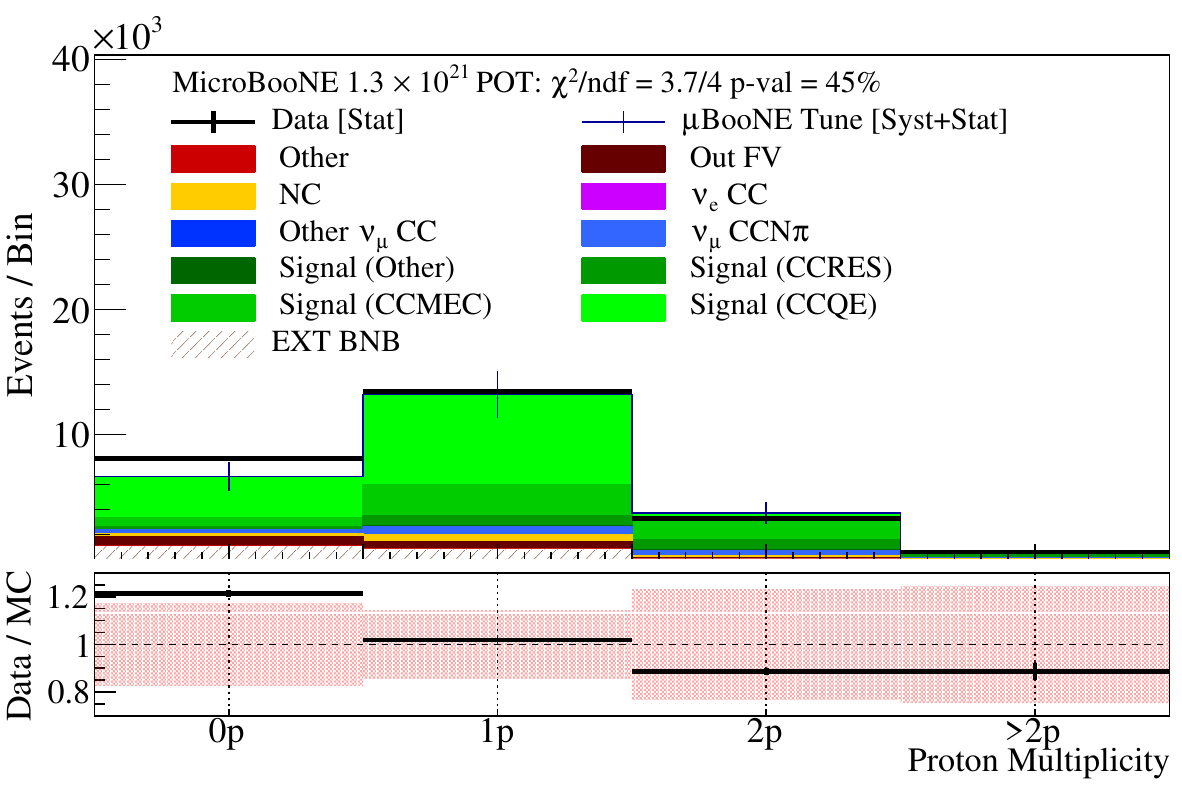}}
\caption{Event rates for the CC$0\pi$ signal selection normalized by bin width,
comparing data to simulation in 1D reconstructed observable bins. The data
points indicate statistical errors, while the errors on the MC represent
systematic uncertainties.}
\label{fig:1d-event-rates}
\end{figure}

\begin{figure*}
\centering
\includegraphics[width=0.9\textwidth]{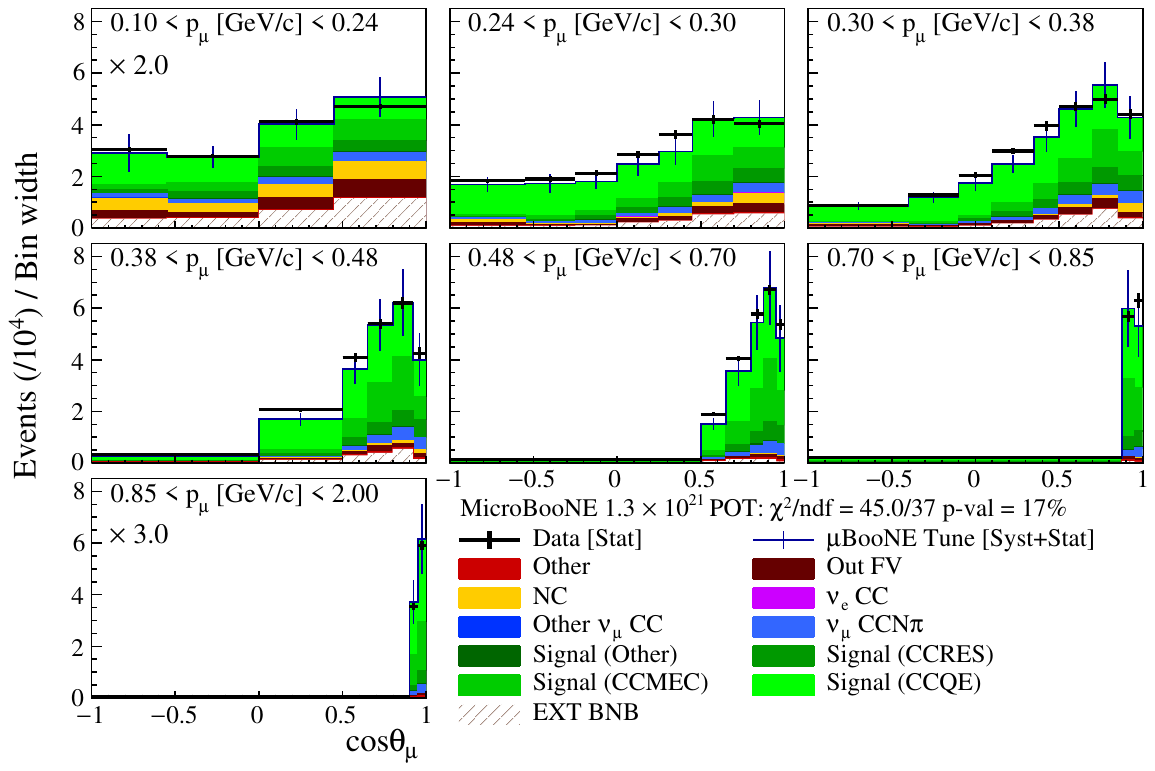}
\caption{Selected CC$0\pi$ events in the 2D reconstructed observable space,
shown as a function of $\cos\theta_\mu$ in regions of $p_\mu$. The data points
indicate statistical errors, while the errors on the MC represent systematic
uncertainties.}
\label{fig:2Dselectioncomparison}
\end{figure*}

Detector response systematics quantify uncertainties arising from discrepancies
between the detector response in simulation and measurements in calibration
control regions~\cite{MicroBooNE:2019efx}. These are estimated using alternative
Monte Carlo samples with modified detector response models, including charge
effects in the LArTPC \cite{MicroBooNE:2021roa, MicroBooNE:2019koz,
MicroBooNE:2020kca}, light propagation and detection, and modified wire signal
formation~\cite{MicroBooNE:2018swd, MicroBooNE:2018vro}. Most detector effects
are captured through a data-driven approach of ``wire
modification''~\cite{MicroBooNE:2021roa}. This technique uses data-simulation
differences in deconvolved wire waveforms for well-characterized, time-tagged
cosmic-ray muon samples to correct the detector's ionization charge response as
a function of position and track angle and pitch. The correction carries an
associated uncertainty driven by the sample
statistics~\cite{MicroBooNE:2021roa}. This method accounts for any source of
ionization charge loss in the TPC. An additional small uncertainty is considered
accounting for the temperature-dependent number of argon target nuclei contained
in the active TPC volume. Detector response uncertainties vary across the
momentum range, as the dominant contribution (32\% of 46\% total) in the lowest
momentum bin. They are subdominant at intermediate and high momentum, with a
minimum of 3\% at $0.5$\,GeV/$c$.

Finally, several statistical uncertainties are considered, accounting for the
finite sample sizes of the main neutrino interaction Monte Carlo, the beam-off
data used to estimate backgrounds due to misidentified cosmic-ray activity in
the absence of a neutrino interaction, and beam-off data samples overlaid with
simulated neutrino interactions. These contributions are in the (1--3)\% range
at $0.5$\,GeV/$c$. Apart from a 9\% contribution in the lowest momentum bin, the
data statistical uncertainties vary from (3--6)\% over the momentum range.

\section{Event rates}
\label{sec:rates}
The measured event rates in 1D and 2D are shown in
Figs.~\ref{fig:1d-event-rates} and \ref{fig:2Dselectioncomparison},
respectively, comparing the full MicroBooNE BNB dataset to the central value
Monte Carlo model (Sec.~\ref{sec:experiment}) with the complete set of
uncertainties (Sec.~\ref{sec:uncert}). Figure~\subref*{fig:1d-event-rates:np}
also shows the reconstructed proton multiplicity distribution for the selected
events. The dominant backgrounds vary across the kinematic space. Overall, the
largest beam neutrino-induced background (5.5\% of the overall selected event
rate) is due to $\nu_\mu$CC$N\pi^\pm$ interactions, where the charged pion was
misidentified. At low $p_\mu$, a larger fraction of background is due to NC
events with final-state protons. Cosmic-ray--related backgrounds (EXT BNB)
constitute 7.8\% of selected events. This is due to the selection's inclusion of
single-track $1\mu0p$ final states, as described in
Sec.~\ref{sec:eventreconstuction}.

\begin{figure}[p!]
\subfloat[Leading proton momentum distribution for the NC sideband selection.
\label{fig:1d-sideband-rates:nc:pp}]{%
\includegraphics[width=0.48\textwidth]{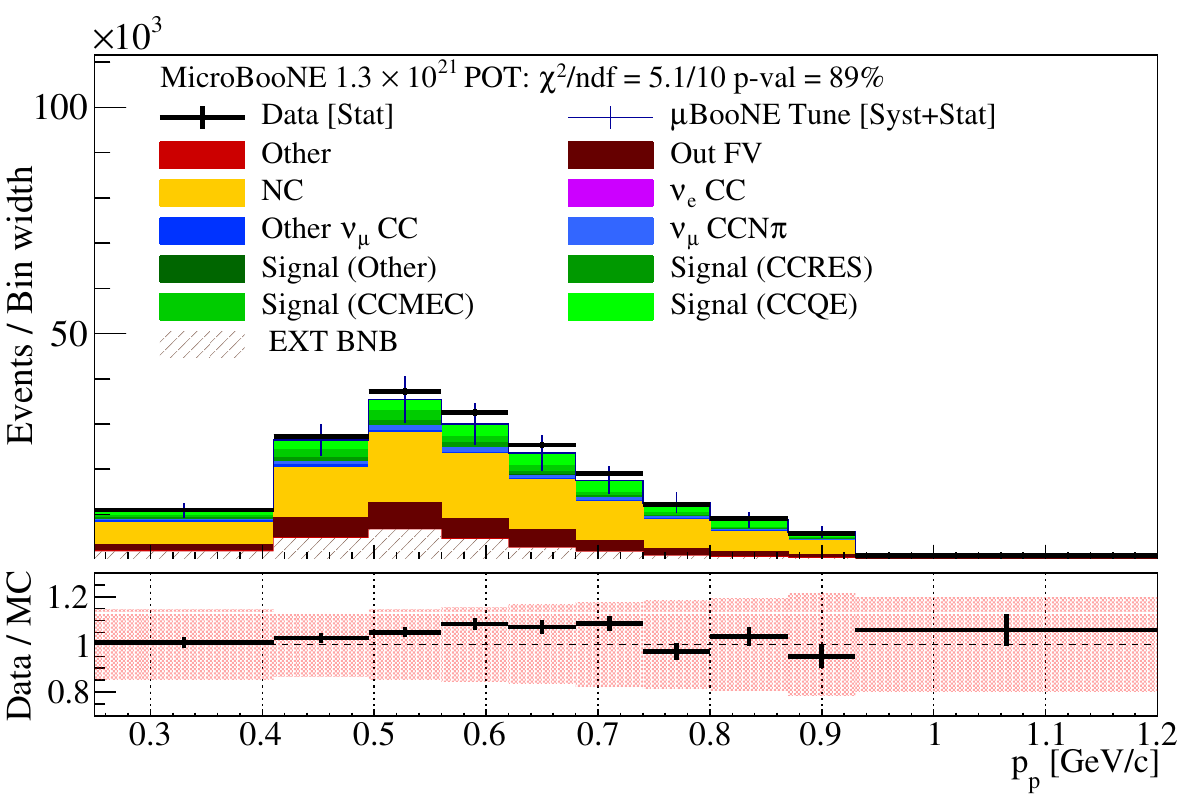}}\\
\subfloat[Distribution of $p_\mu$ for the CC$N\pi$ sideband selection.
\label{fig:1d-sideband-rates:npi:pmu}]{%
\includegraphics[width=0.48\textwidth]{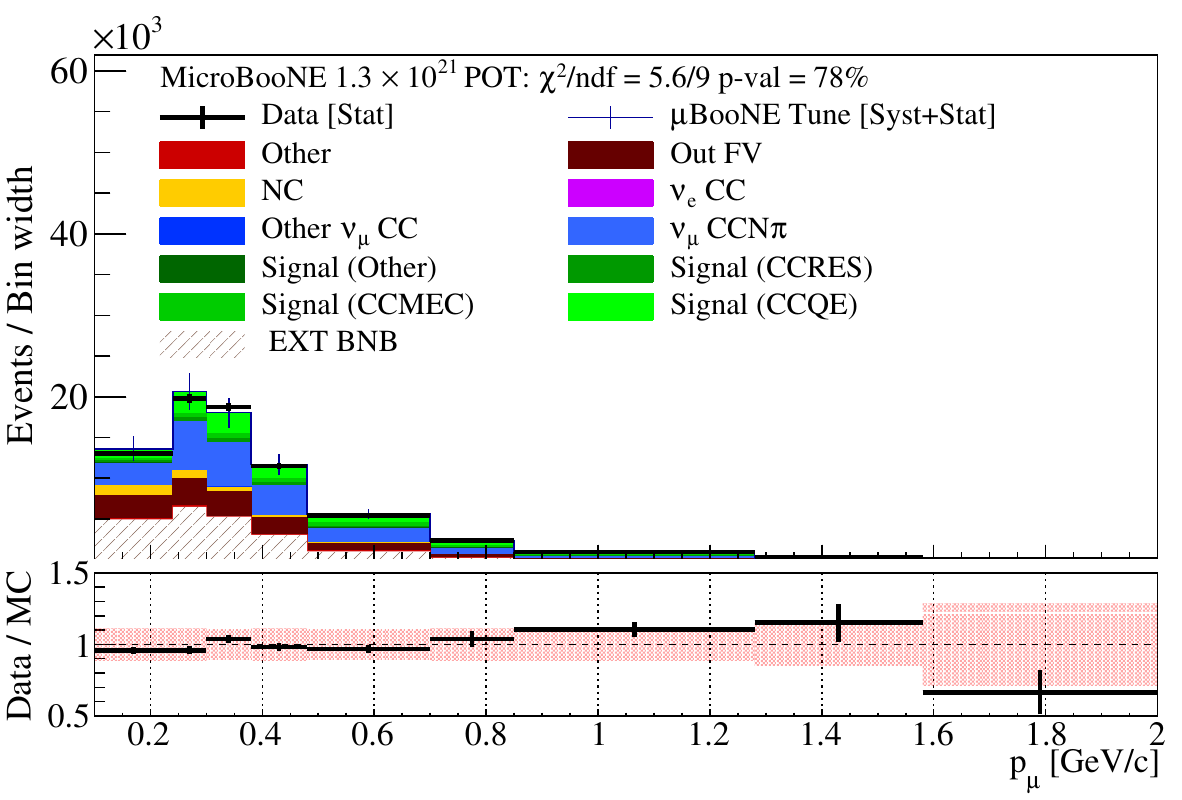}}\\
\subfloat[Distribution of $\cos\theta_\mu$ for the CC$N\pi$ sideband selection.
\label{fig:1d-sideband-rates:npi:ctmu}]{%
\includegraphics[width=0.48\textwidth]{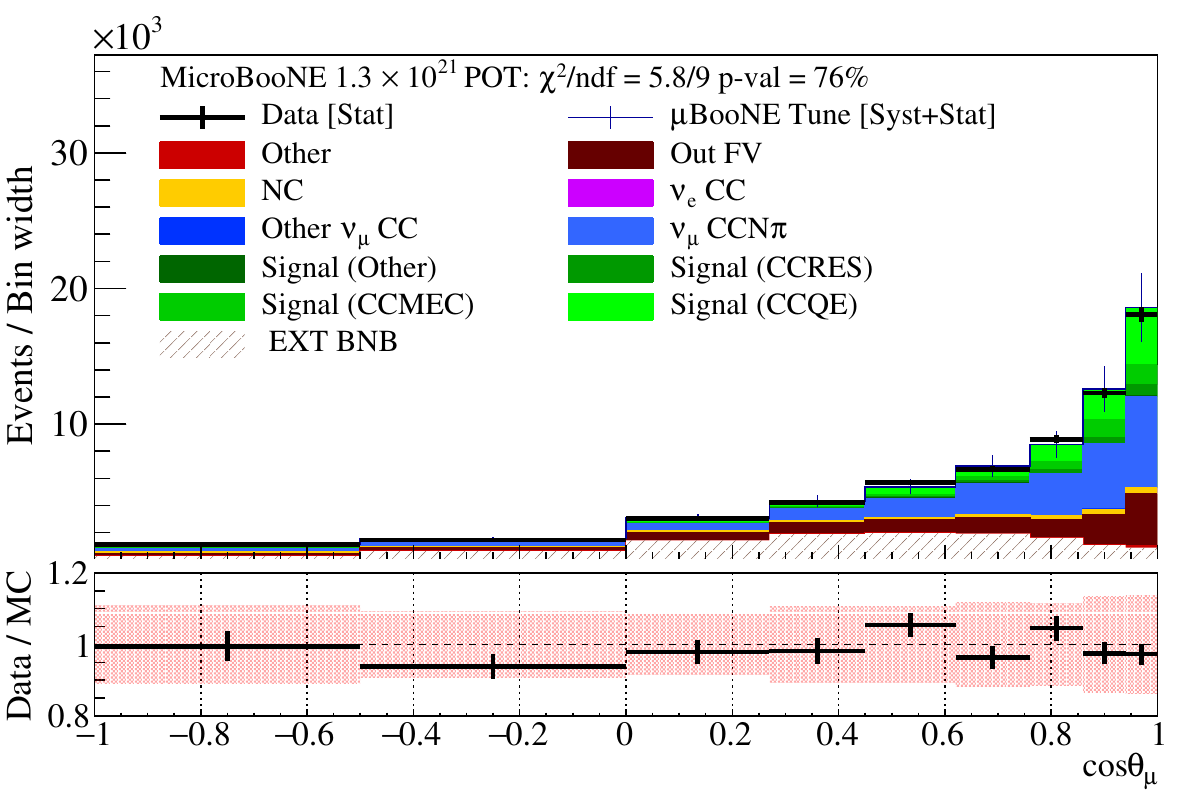}}
\caption{Event rates for the validation sidebands, comparing data to simulation.
The data points indicate statistical errors, while the errors on the MC
represent systematic uncertainties.}
\label{fig:1d-sideband-rates}
\end{figure}

To validate the background modeling in the signal region, we consider a set of
sideband control regions, defined by inverting selection criteria. The
NC sideband selection requires that there are no muonlike particles
in the final state. Here, events are selected having at least one protonlike
rather than muonlike final-state track. The BDT PID model must identify at
least one proton in the final state, and all protons are required to have a
momentum in the range $(0.25-1.0)$\,GeV/$c$.
Figure~\subref*{fig:1d-sideband-rates:nc:pp} shows the event rate for the NC
sideband as a function of the leading proton momentum. We obtain a sample with
approximately 45\% purity in NC events, the CC$0\pi$ signal being strongly
suppressed, and find good agreement ($p=89$\%) within uncertainties. The total
uncertainty of about 25\% is dominated by interaction modeling uncertainties.

A CC$N\pi$ sideband provides a validation of the modeling of pion backgrounds in
the CC$0\pi$ signal region. Here, events are initially classified as zero-shower
$\nu_\mu$CC-like and must satisfy the preselection cuts outlined in
Sec.~\ref{sec:preselection}. From this sample, events are selected where the
BDT PID model identifies one or more charged pionlike final-state tracks.
Candidate pion tracks must exceed a momentum threshold of 70\,MeV/$c$, motivated
by the track reconstruction efficiency.
Figures~\subref*{fig:1d-sideband-rates:npi:pmu} and
\subref*{fig:1d-sideband-rates:npi:ctmu} show the resulting CC$N\pi$ event
distribution, measured as a function of the final-state reconstructed $p_\mu$
and $\cos\theta_{\mu}$. Again in this selection, enhanced to 27.9\% purity in
$\nu_\mu$CC$N\pi$ events, we observed a high level of compatibility ($p>75$\%)
with the central value MicroBooNE GENIE tune simulation. A final sideband
inverts the fiducial volume cut to measure the rates of events near the detector
boundary, which contribute to the out-of-fiducial-volume backgrounds.
This sample also shows agreement between data and simulation with $p=85$\% for
the reconstructed muon momentum distribution. On the basis that the NC,
CC$N\pi$, and out-of-fiducial-volume backgrounds appear to be well modeled within our
uncertainties, we do not apply any additional tuning or corrections to the
corresponding backgrounds in the signal region.

\section{Cross-section extraction}
\label{sec:sigex}

We subtract estimated backgrounds from the event rates measured in data, scale
according to the integrated neutrino flux and number of target nuclei, and apply
efficiency and acceptance corrections to account for bin migrations between true
and reconstructed observable bins. The full set of correlated systematic
uncertainties as expressed in a covariance matrix are propagated. These
corrections, which map the observed event rates from the space of reconstructed
observables to a space of regularized true observables, are applied through an
unfolding procedure. For this measurement, the unfolding is performed using the
Wiener-SVD method~\cite{Tang:2017rob} using the first derivative to define the
regularization penalty term~\cite{Gardiner:2024gdy}. This technique provides a
regularization that preserves the statistical power of goodness-of-fit tests
relative to the directly measured reconstructed distributions. This unfolding
approach is described in further detail in Ref.~\cite{MicroBooNE:2024zwf}.

The degree of regularization applied in unfolding using the Wiener-SVD method is
represented in an output matrix referred to as the regularization matrix
$A_{C}$. This matrix encodes the transformation between true observables and the
regularized observables we use to express our measurement. Distributions
expressed in terms of true observables (e.g., generator $p_\mu$) must be
transformed by the $A_C$ matrix for statistical comparisons to this data.

Prior to analysis of the full dataset, a series of tests were conducted using
Monte Carlo samples treated as data to validate the unfolding procedure and
assess potential model dependence. The NuWro generator~\cite{Golan:2012rfa}
provides an alternative neutrino interaction model. Additional samples were
generated with extreme variations on the base GENIE model: one disabling the
MicroBooNE GENIE tuning~\cite{MicroBooNE:2021ccs} (see
Sec.~\ref{sec:microboone:basemodel}) and another increasing the MEC ($2p2h$)
cross section by a factor of 2. In each case, we extract the unfolded cross
section and assess the consistency with the true underlying cross section under
the corresponding model variation hypothesis, determining the statistical
consistency using a $\chi^2$ test. As the flux and detector simulation are
identical, the only uncertainties included are those on the cross-section model
and, additionally, the Monte Carlo statistical uncertainties for the NuWro
comparisons. The extracted cross sections are found to be well within the
uncertainties in all cases, with $p>73$\% for the NuWro simulation and $p>84$\%
for the alternative GENIE models.  As an additional test of the uncertainties
impacting momentum scale, a set of simulations are constructed by shifting the
true and reconstructed muon momenta upward by 5\% and 10\%, about equal to and
twice the muon momentum resolution, respectively. For these tests, the full set
of uncertainties is used. With $p$~values of 94\% and 41\% for the extracted
cross sections relative to the true cross sections, we observe that the
unfolding procedure is capable of recovering the true cross section even in the
case of a significant mismodeling of the muon momentum scale well beyond the
relevant uncertainties.

\section{Results}
\label{sec:results}
\begin{figure*}
\centering
\subfloat[$p_\mu$ differential cross section.
\label{fig:xs-1d:pmu}]{%
\includegraphics[width=0.65\textwidth]{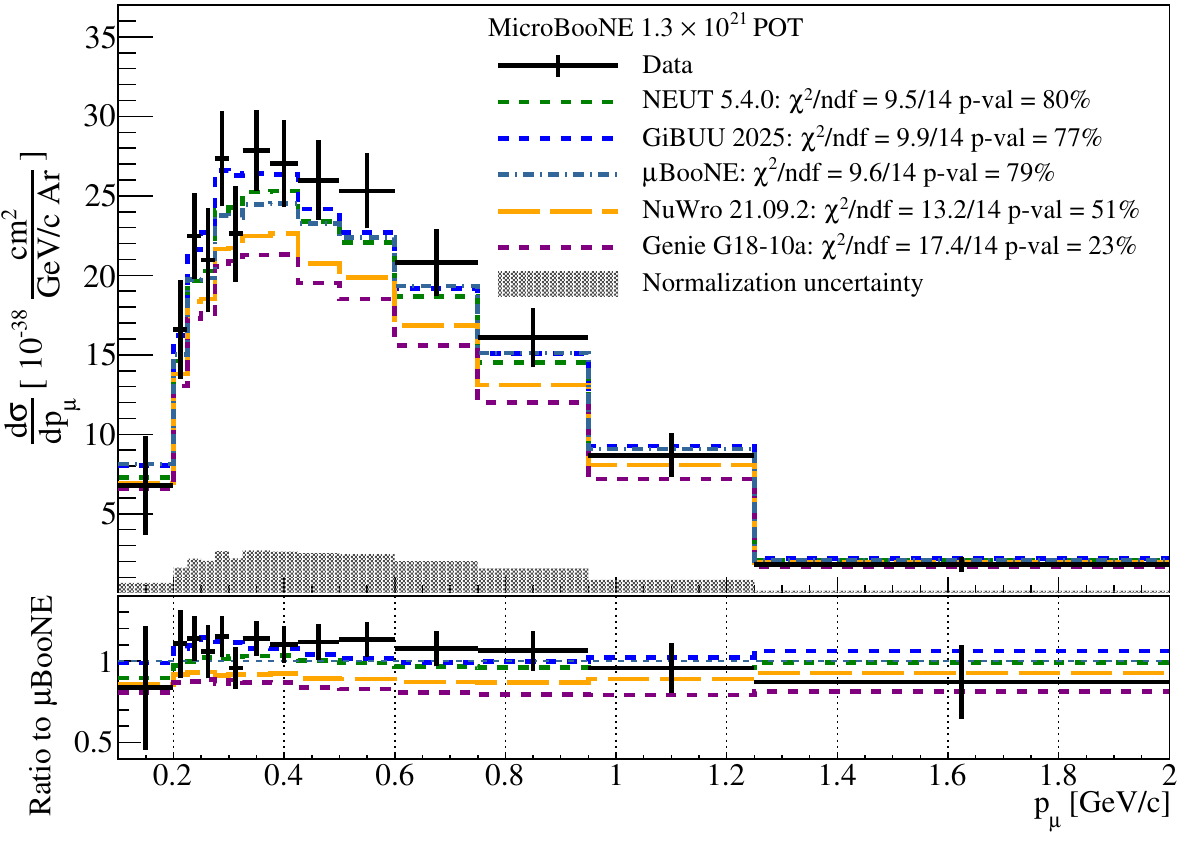}}\\
\subfloat[$\cos\theta_\mu$ differential cross section.
\label{fig:xs-1d:ctmu}]{%
\includegraphics[width=0.65\textwidth]{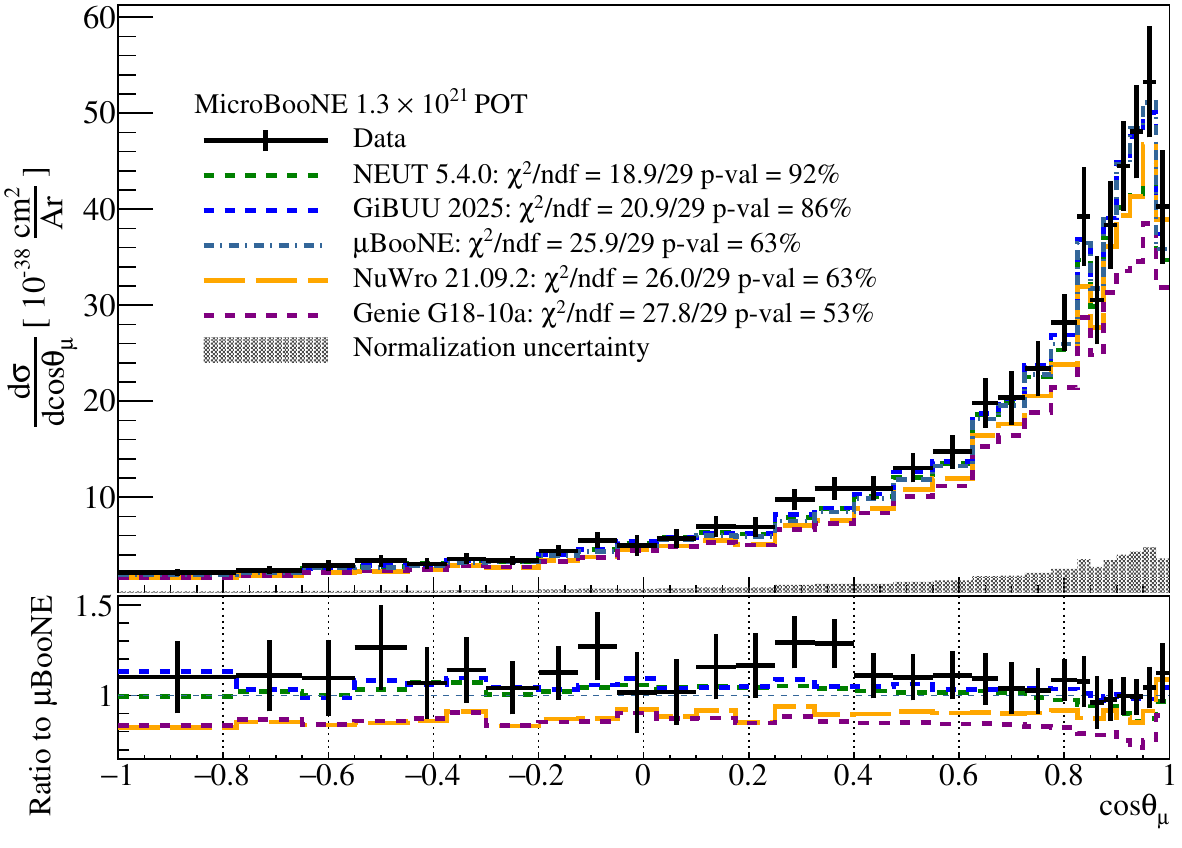}}\\
\caption{Extracted flux-integrated differential cross sections for the
CC$0\pi$ signal, in 1D $p_\mu$ and $\cos\theta_\mu$ bins. Errors on the data
include both statistical and systematic uncertainties.}
\label{fig:xs-1d}
\end{figure*}

\begin{figure*}
\includegraphics[width=0.42\textwidth,page=2]{2DSlices_crossSection.pdf}
\includegraphics[width=0.42\textwidth,page=3]{2DSlices_crossSection.pdf}
\includegraphics[width=0.42\textwidth,page=4]{2DSlices_crossSection.pdf}
\includegraphics[width=0.42\textwidth,page=5]{2DSlices_crossSection.pdf}
\includegraphics[width=0.42\textwidth,page=6]{2DSlices_crossSection.pdf}
\includegraphics[width=0.42\textwidth,page=7]{2DSlices_crossSection.pdf}
\includegraphics[width=0.42\textwidth,page=8]{2DSlices_crossSection.pdf}
\includegraphics[width=0.42\textwidth,page=1]{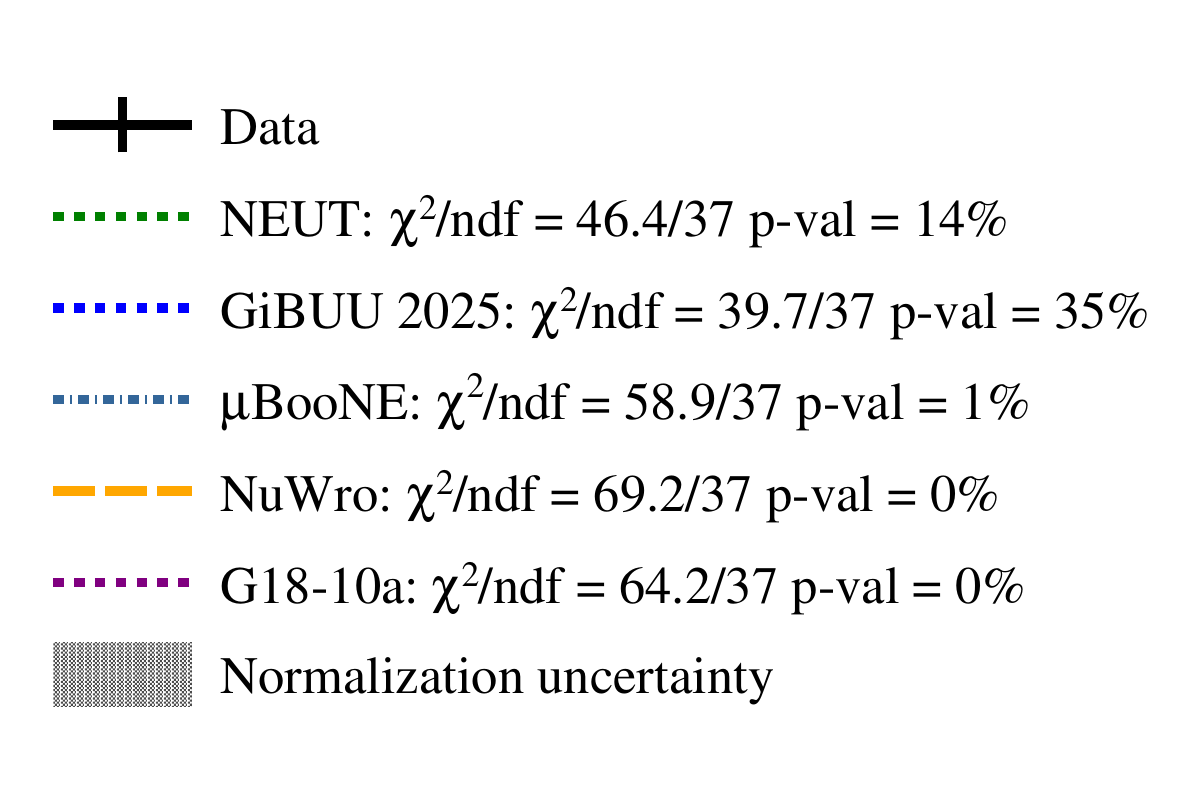}
\caption{Extracted flux-integrated double-differential cross sections for the
CC$0\pi$ signal, shown as a function of $\cos\theta_\mu$ in ranges of $p_\mu$.
Errors on the data include both statistical and systematic uncertainties.
The $\chi^2$ and $p$~values for individual momentum ranges are listed in
Table~\ref{tab:chi2_slices}.}
\label{fig:xs-2d}
\end{figure*}

The unfolded single-differential cross sections in $p_\mu$ and $\cos\theta_\mu$
are shown in Figs.~\subref*{fig:xs-1d:pmu} and \subref*{fig:xs-1d:ctmu}. The
double-differential cross section is shown in Fig.~\ref{fig:xs-2d} as a function
of $\cos\theta_\mu$ in ranges of $p_\mu$. For the double-differential cross
section, the $\chi^2$ values and $p$ values in each momentum range are given in
Table~\ref{tab:chi2_slices}. The fractional uncertainties on the extracted
cross sections are available in Supplemental Material~\cite{supplement}. We
compare the extracted cross sections to predictions from a variety of neutrino
event generators, transformed to the unfolded observable space via application
of the $A_C$ matrix. The corresponding $\chi^2$ statistic and $p$~value relative
to the unfolded data are also shown.

\begin{table*}
\caption{Generator model comparisons for individual momentum ranges in the
double-differential cross-section measurement, indicating the $\chi^2$ and
$p$~values ($p$) for each generator in each range. Momentum ranges are in units
of GeV/$c$.}
\label{tab:chi2_slices}
\centering
\renewcommand{\arraystretch}{1.2}
\begin{tabular}{|l|c c|c c|c c|c c|c c|c c|c c|}
\hline
\multirow{2}{17ex}{\diagbox[height=2.4\line, width=18ex]{Model}{$p_\mu$ range}} &
\multicolumn{2}{c|}{0.10 -- 0.24} & \multicolumn{2}{c|}{0.24 -- 0.30} &
\multicolumn{2}{c|}{0.30 -- 0.38} & \multicolumn{2}{c|}{0.38 -- 0.48} &
\multicolumn{2}{c|}{0.48 -- 0.70} & \multicolumn{2}{c|}{0.70 -- 0.85} &
\multicolumn{2}{c|}{0.85 -- 2.00} \\  
\cline{2-15} &
$\chi^2/\text{n.d.f.}$ & $p$ & $\chi^2/\text{n.d.f.}$ & $p$ &
$\chi^2/\text{n.d.f.}$ & $p$ & $\chi^2/\text{n.d.f.}$ & $p$ &
$\chi^2/\text{n.d.f.}$ & $p$ & $\chi^2/\text{n.d.f.}$ & $p$ &
$\chi^2/\text{n.d.f.}$ & $p$ \\
\hline
$\mu$BooNE & 1.1/4 & 89\% &  5.4/7 & 61\% &  14.0/8 & 8\%  & 17.0/6 &  1\% &  6.7/6 & 35\% &  3.3/3 & 35\% & 0.4/3 & 95\%\\\hline
NEUT       & 2.5/4 & 65\% &  2.6/7 & 92\% &  5.8/8  & 67\% &  7.0/6 & 32\% & 11.2/6 &  8\% &  5.8/3 & 12\% & 2.4/3 & 49\%\\\hline
GiBUU 2025 & 1.3/4 & 85\% &  1.4/7 & 99\% &  4.6/8  & 80\% &  6.4/6 & 38\% &  4.1/6 & 67\% &  4.7/3 & 20\% & 0.2/3 & 98\%\\\hline
NuWro      & 4.3/4 & 37\% & 11.7/7 & 11\% & 18.4/8  & 2\%  & 32.7/6 &  0\% & 16.4/6 &  1\% &  5.6/3 & 14\% & 2.6/3 & 46\%\\\hline
G18-10a    & 5.8/4 & 22\% &  8.0/7 & 21\% & 14.1/8  &  6\% & 24.6/6 &  0\% & 22.9/6 &  0\% & 12.3/3 &  1\% & 6.9/3 &  8\%\\\hline
\end{tabular}
\end{table*}

The measurements are compared to models including: (i) {GENIE} version
v3.2.0~\cite{GENIE:2021npt} using the {G18\_10a\_02\_11a} CMC (ii)
{GiBUU 2025}~\cite{Buss:2011mx}; (iii) {NEUT} version
5.4.0.1~\cite{Hayato:2021heg}; and (iv) {NuWro} version
21.09.2~\cite{Golan:2012rfa}. These generator predictions are processed using
the {NUISANCE} framework~\cite{Stowell:2016jfr}. Table
\ref{tab:chi2_combined_full} summarizes the resulting $\chi^{2}$ and $p$~values
comparing the measured data with the models. A detailed comparison of the models
represented in each generator configuration, as well as results for additional
generator models, are available in Supplemental
Material~\cite{supplement}.

We find that all models considered here describe the data in individual 1D
distributions, while only a subset agree with the double-differential
measurement. The latter provides a stricter test, assessing the extent to which
models correctly describe the angular distribution as a function of momentum
and considering the correlations in this 2D space. Overall, {GiBUU 2025}
performs best for the measurement across the fully correlated 2D space, and the
{NEUT} model also performs well. The success of up-to-date neutrino event
generators is indicative of the significant modeling improvements in recent
years and the value of multidimensional measurements to assess models. Most
generators underpredict the normalization. This effect is most pronounced for
forward angles and intermediate muon momentum $(0.5-0.7)$\,GeV/$c$. These
effects can be seen in Fig.~\ref{fig:xs-2d}, which indicates the goodness of fit
for the angular distribution within each momentum range individually.

\begin{table*}
\caption{Summary of generator model comparisons, indicating the $\chi^2$ and
$p$~values for each model considered, in comparison to the 2D extracted cross section and the individual 1D single-differential distributions in $p_\mu$ and
$\cos\theta_\mu$.}
\label{tab:chi2_combined_full}
\centering
\renewcommand{\arraystretch}{1.2}
\setlength{\tabcolsep}{6pt}
\begin{tabular}{|l|c c|c c|c c|}
\hline
\multirow{2}{17ex}{{Model/Tune}} &
\multicolumn{2}{c|}{2D $p_{\mu}/\cos\theta_\mu$} &
\multicolumn{2}{c|}{$p_{\mu}$} &
\multicolumn{2}{c|}{$\cos\theta_{\mu}$} \\  
\cline{2-7} &
$\chi^2/\text{n.d.f.}$ &
$p$~value &
$\chi^2/\text{n.d.f.}$ &
$p$~value &
$\chi^2/\text{n.d.f.}$ &
$p$~value \\  
\hline
\hline
$\mu$BooNE & 58.9/37 &  1\% &  9.6/14 & 79\% & 25.9/29 & 63\% \\
\hline
NEUT       & 46.4/37 & 14\% &  9.5/14 & 80\% & 18.9/29 & 92\% \\
\hline
GiBUU 2025 & 39.7/37 & 35\% &  9.9/14 & 77\% & 20.9/29 & 86\% \\
\hline
NuWro      & 69.2/37 &  0\% & 13.2/14 & 51\% & 26.0/29 & 63\% \\
\hline
G18-10a    & 64.2/37 &  0\% & 17.4/14 & 23\% & 27.8/29 & 53\% \\
\hline
\end{tabular}
\end{table*}

\section{Conclusions}
\label{sec:conclusions}
Neutrino interaction measurements using LArTPCs provide crucial input to inform
the development and tuning of the models underpinning precision neutrino
oscillation measurements and background estimation for beyond the Standard Model
physics searches. Using the full $1.3\times10^{21}$\,POT exposure of the
MicroBooNE LArTPC experiment in the Fermilab BNB, we have performed a
high-statistics measurement of the flux-integrated single- and
double-differential cross sections in terms of final-state muon kinematics for
$\nu_\mu$CC$0\pi$ interactions on argon, i.e., interactions that have no
final-state pions but may or may not have final-state protons. This channel is
of importance for neutrino oscillation searches as a topologically defined proxy
for well-modeled CCQE interactions, especially in detectors with a high
threshold for hadronic activity (e.g., water Cherenkov experiments), for which this
measurement with an argon target provides a point of comparison.

The measurement, unfolding techniques, and background estimation are validated
using a suite of fake data tests and targeted sideband selections isolating NC
and CC$N\pi^\pm$ interactions. The observed event rates show good agreement with
the central value GENIE model used in the efficiency and background estimation.
Considering a suite of uncertainties concerning the data statistics, neutrino
flux, interaction model, detector response, cosmic-ray background normalization,
and exposure and target mass estimation, we find an average total uncertainty
below 20\%.

The extracted 1D and 2D cross sections as a function of $p_\mu$ and
$\cos\theta_\mu$ are quantitatively compared with a variety of neutrino event
generators, representing an array of physics model sets and implementations. We
find that all generator models considered provide a reasonable description of
the $p_\mu$ and $\cos\theta_\mu$ distributions individually. The {GiBUU 2025} and {NEUT} models also show agreement with the correlated joint
distribution, while other models yield a poorer description with $p$ values
$<5$\%. Additional tests using alternative GENIE models provide further guidance
on the impact of specific model implementations in regions of the kinematic
phase space. These data provide a high-statistics measurement of
$\nu_\mu$CC$0\pi$ interactions in the correlated joint space of muon momentum
and angle and offer a benchmark for further modeling improvements, looking ahead
to increasingly precise measurements of neutrino oscillations and other
accelerator-based measurements using both argon and other nuclei.

\begin{acknowledgments}
This document was prepared by the MicroBooNE Collaboration using the resources
of the Fermi National Accelerator Laboratory (Fermilab), a U.S. Department of
Energy, Office of Science, Office of High Energy Physics HEP user facility.
Fermilab is managed by Fermi Forward Discovery Group, LLC, acting under Contract
No. 89243024CSC000002.  MicroBooNE is supported by the following: the U.S.
Department of Energy, Office of Science, Offices of High Energy Physics and
Nuclear Physics; the U.S. National Science Foundation; the Swiss National
Science Foundation; the Science and Technology Facilities Council (STFC), part
of the United Kingdom Research and Innovation; the Royal Society (United
Kingdom); the United Kingdom Research and Innovation (UKRI) Future Leaders Fellowship; and
the NSF AI Institute for Artificial Intelligence and Fundamental Interactions.
Additional support for the laser calibration system and cosmic-ray tagger was
provided by the Albert Einstein Center for Fundamental Physics, Bern,
Switzerland. We also acknowledge the contributions of technical and scientific
staff to the design, construction, and operation of the MicroBooNE detector as
well as the contributions of past collaborators to the development of MicroBooNE
analyses, without whom this work would not have been possible.
\end{acknowledgments}

\section*{Data availability}

The data that support the findings of this article are not publicly available. The data are available from the authors
upon reasonable request.

\bibliography{references}

\end{document}


\title{Supplemental Material: Measurement of charged-current muon
neutrino--argon interactions without pions in the final state using the
MicroBooNE detector}

\begin{abstract}
This supplemental material contains additional details regarding the measurement
presented in the main text. This includes specifics on the binning of
observables, bin migration (i.e., smearing and acceptance), breakdowns of the
fractional uncertainties on each cross-section bin, the additional smearing
matrices associated with the unfolding that are required for comparisons to the
measured data, and details on generator model comparisons.
\end{abstract}

\maketitle

\section{Binning}

This section provides the bin edges for the 1D and 2D differential measurements
in CC$0\pi$ muon kinematics, as obtained using the procedure described in the
main text.

\subsection{Single-differential}

For the 1D single-differential cross sections, we define and optimize bins
separately in $p_{\mu}$ and $\cos\theta_\mu$:

\begin{equation}
\begin{split}
p_{\mu}~[\mathrm{GeV}/c]:~
(&0.1,   ~0.2,  ~0.225, ~0.25, \\
 &0.275, ~0.3,  ~0.325, ~0.375,\\
 &0.425, ~0.5,  ~0.6,   ~0.75, \\
 &0.95,  ~1.25, ~2),
\end{split}
\end{equation}

\begin{equation}
\begin{split}
\cos\theta_\mu:~
(& -1,   ~-0.775, ~-0.65, ~-0.55,  ~-0.45,\\
&-0.375, ~-0.3,   ~-0.2,  ~-0.125, ~-0.05,\\
&0.025,  ~0.1,    ~0.175, ~0.25,   ~0.325,\\
&0.4,    ~0.475,  ~0.55,  ~0.625,  ~0.675,\\
&0.725,  ~0.775,  ~0.825, ~0.85,   ~0.875,\\
&0.9,    ~0.925,  ~0.95,  ~0.975,  ~1).
\end{split}
\end{equation}

\subsection{Double-differential}

In the 2D binning, we define slices of $p_{\mu}$ as a function of
$\cos\theta_{\mu}$, with the bin edges listed in Table~\ref{tab:bins-2d}.

\begin{table}[h!]
\caption{Edges for 2D binning. Each section contains the $\cos\theta_\mu$ edges for
the a given $p_\mu$ bin range. Momenta are in units of GeV/$c$.}
\label{tab:bins-2d}
\label{table:2D_blockBinning_scheme1}
\centering
\begin{tabular}{|c|cc|cc|c|}
\hline
ID
& $p_\mu^\mathrm{low}$
& $p_\mu^\mathrm{high}$
& $\cos\theta_\mu^\mathrm{low}$
& $\cos\theta_\mu^\mathrm{high}$
& Efficiency \\
\hline
\hline
1 & 0.1 & 0.24 & -1 & -0.55 & 0.23 \\
2 &  & & -0.55 & 0 & 0.22 \\
3 &  & & 0 & 0.45 & 0.26 \\
4 &  & & 0.45 & 1 & 0.29 \\
\hline
5 & 0.24 & 0.3 & -1 & -0.55 & 0.27 \\
6 &  & & -0.55 & -0.25 & 0.24 \\
7 &  & & -0.25 & 0 & 0.23 \\
8 &  & & 0 & 0.25 & 0.26 \\
9 &  & & 0.25 & 0.45 & 0.26 \\
10 &  & & 0.45 & 0.7 & 0.33 \\
11 &  & & 0.7 & 1 & 0.38 \\
\hline
12 & 0.3 & 0.38 & -1 & -0.4 & 0.19 \\
13 &  & & -0.4 & -0.1 & 0.15 \\
14 &  & & -0.1 & 0.1 & 0.15 \\
15 &  & & 0.1 & 0.35 & 0.22 \\
16 &  & & 0.35 & 0.5 & 0.26 \\
17 &  & & 0.5 & 0.7 & 0.28 \\
18 &  & & 0.7 & 0.85 & 0.33 \\
19 &  & & 0.85 & 1 & 0.38 \\
\hline
20 & 0.38 & 0.48 & -1 & 0 & 0.11 \\
21 &  & & 0 & 0.5 & 0.15 \\
22 &  & & 0.5 & 0.65 & 0.20 \\
23 &  & & 0.65 & 0.8 & 0.27 \\
24 &  & & 0.8 & 0.92 & 0.30 \\
25 &  & & 0.92 & 1 & 0.39 \\
\hline
26 & 0.48 & 0.7 & -1 & 0.5 & 0.06 \\
27 &  & & 0.5 & 0.65 & 0.11 \\
28 &  & & 0.65 & 0.8 & 0.15 \\
29 &  & & 0.8 & 0.875 & 0.20 \\
30 &  & & 0.875 & 0.95 & 0.24 \\
31 &  & & 0.95 & 1 & 0.30 \\
\hline
32 & 0.7 & 0.85 & -1 & 0.875 & 0.06 \\
33 &  & & 0.875 & 0.95 & 0.16 \\
34 &  & & 0.95 & 1 & 0.23 \\
\hline
35 & 0.85 & 2 & -1 & 0.9 & 0.03 \\
36 &  & & 0.9 & 0.95 & 0.07 \\
37 &  & & 0.95 & 1 & 0.10 \\
\hline
\end{tabular}
\end{table}

\section{Bin migration}

\begin{figure}[h!]
\centering
\subfloat[\label{fig:migration:1d:pmu}]{%
\includegraphics[width=0.48\textwidth]{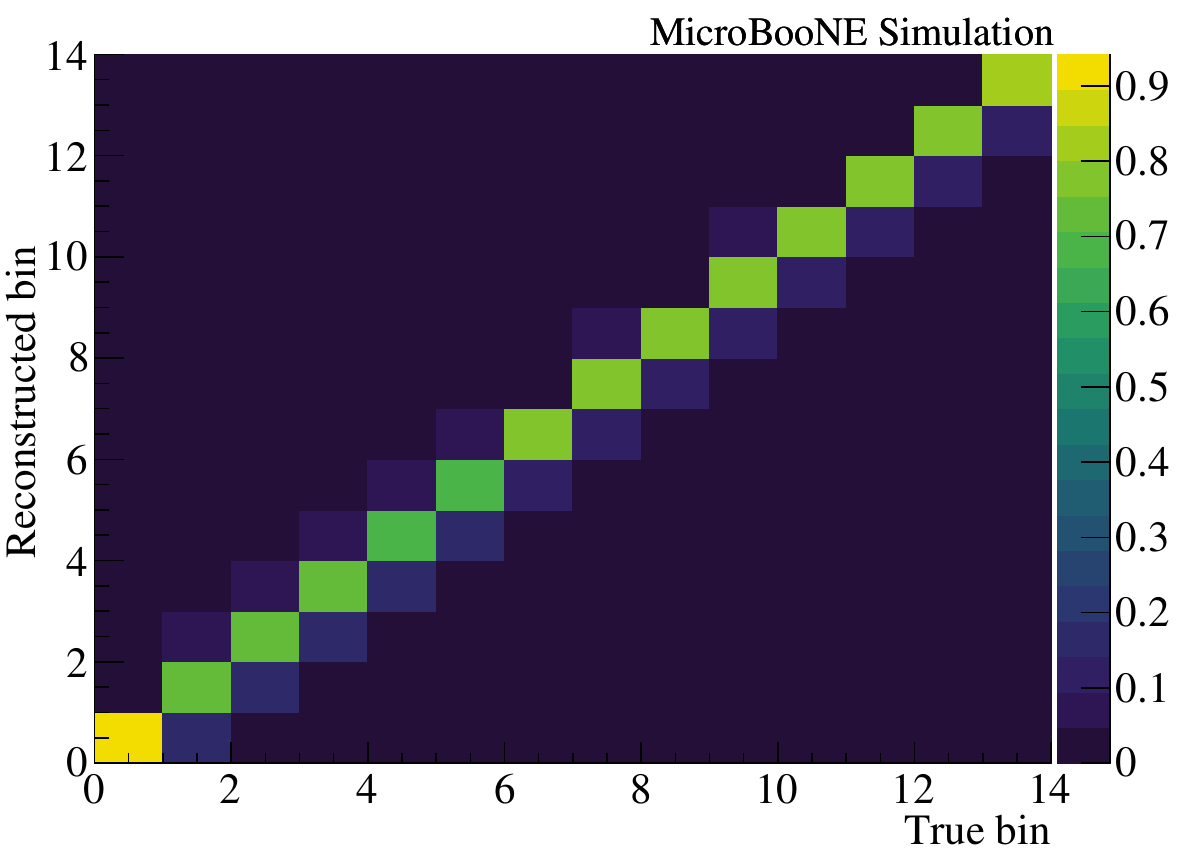}}\\
\subfloat[\label{fig:migration:1d:ctmu}]{%
\includegraphics[width=0.48\textwidth]{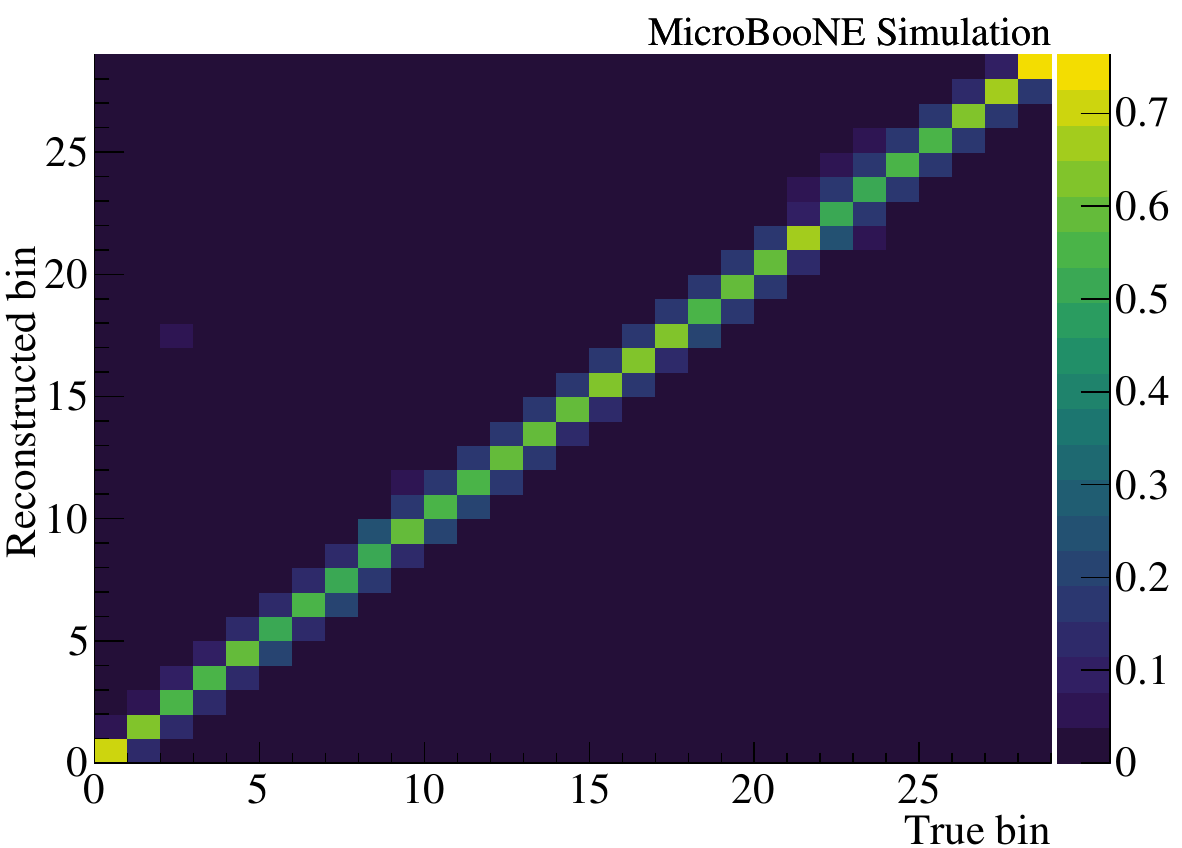}}\\
\subfloat[\label{fig:migration:2d}]{%
\includegraphics[width=0.48\textwidth]{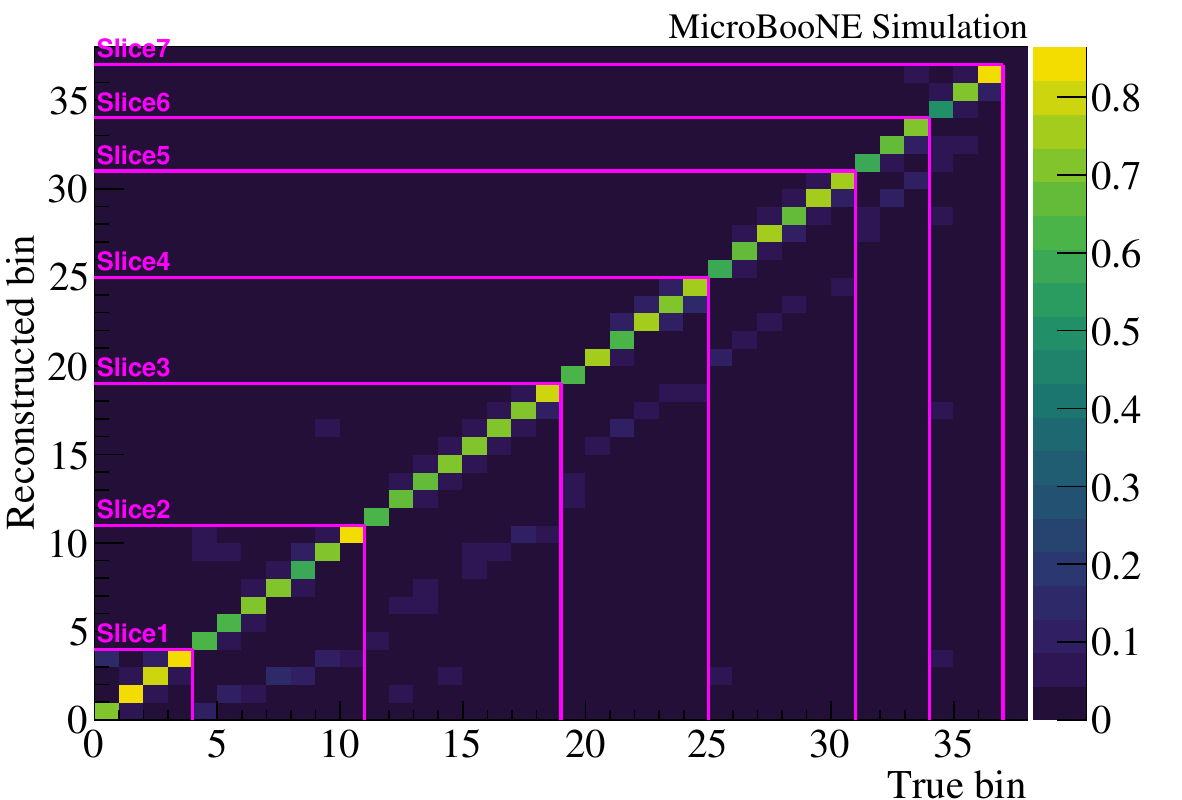}}
\caption{Migration matrices, showing the distribution of events in the space of
reconstructed observables for each true observable bin for (a) 1D $p_\mu$,
(b) 1D $\cos\theta_\mu$, and (c) 2D $p_\mu-\cos\theta_\mu$.}
\label{fig:migration}
\end{figure}

The migration of events between true and reconstructed bins is represented by
the migration (smearing) matrix. Figures~\subref*{fig:migration:1d:pmu} and
\subref*{fig:migration:1d:ctmu} show the bin migration between true and
reconstructed $p_\mu$ and $\cos\theta_\mu$ bins for the individual 1D event rate
distributions. Figure~\subref*{fig:migration:2d} shows the migration matrix for
the 2D event rate distribution, with segments showing the bins of $p_\mu$. The
bin definition procedure requires $\geq50$\% of the events in a given true bin
to fall into the corresponding reconstructed bin, i.e., along the diagonal of
the migration matrix.

\section{Fractional uncertainties}

\begin{figure*}[h!]
\centering
\subfloat[\label{fig:uncert-1d-pmu}]{%
\includegraphics[width=0.5\textwidth]{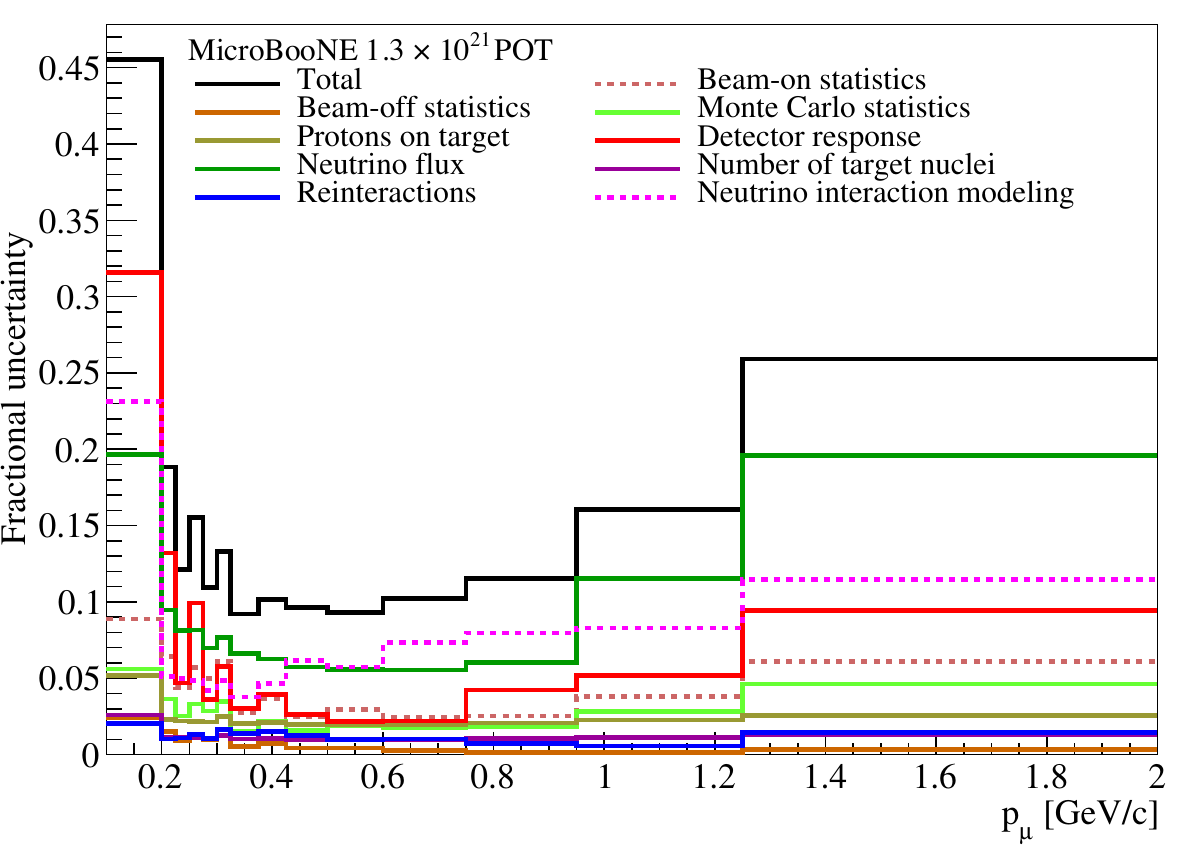}}
\hfill
\subfloat[\label{fig:uncert-1d-ctmu}]{%
\includegraphics[width=0.5\textwidth]{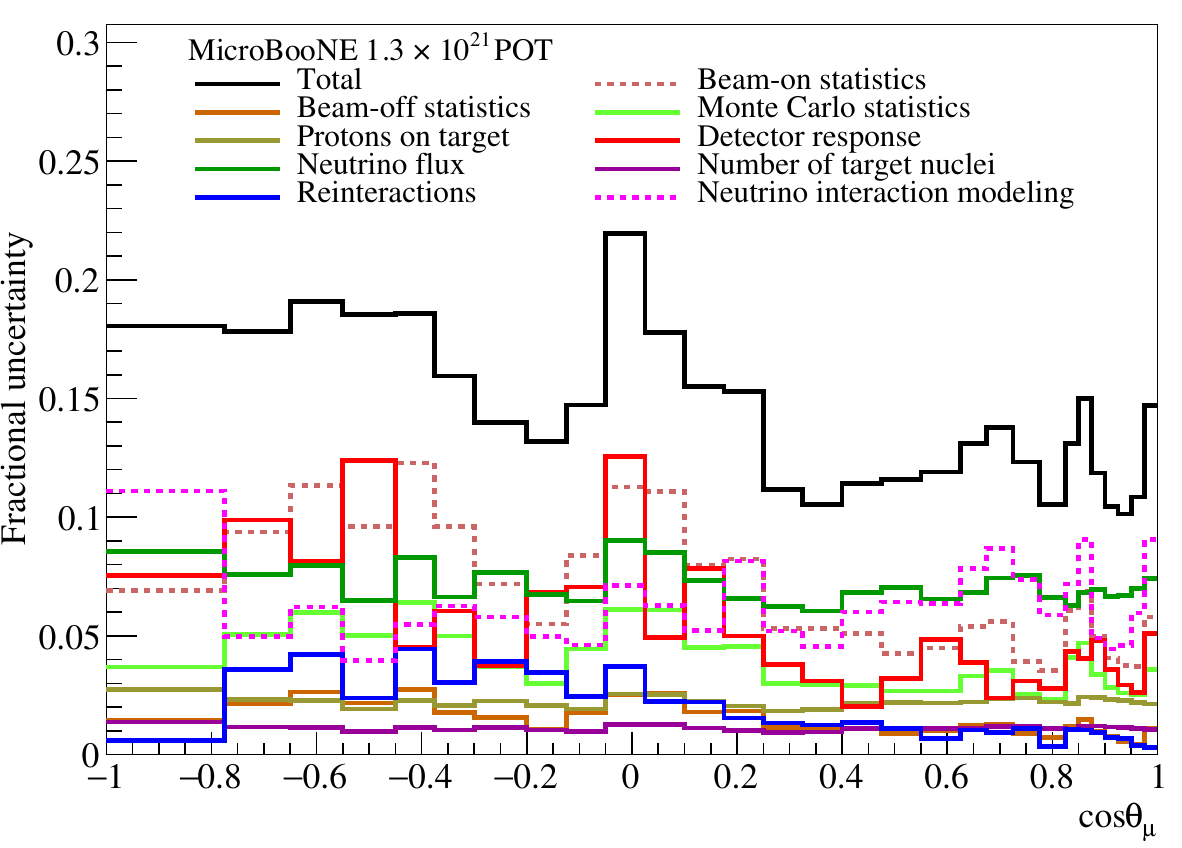}}
\caption{Fractional uncertainties on each bin in the corresponding cross sections for
(a) $p_\mu$ and (b) $\cos\theta_\mu$.}
\label{fig:xs-1d-uncert}
\end{figure*}

\begin{figure*}
\centering
\includegraphics[width=0.8\textwidth]{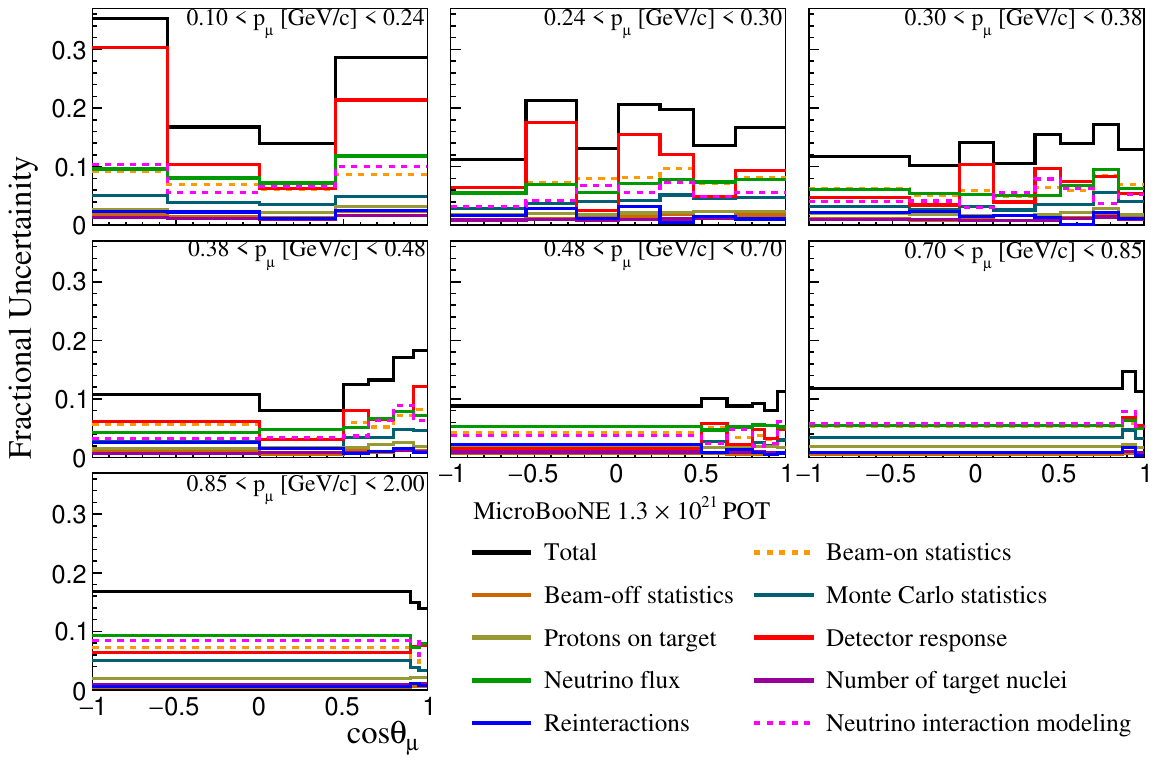}
\caption{Fractional uncertainties for each bin in the double-differential
extracted cross section.}
\label{fig:uncert-2d}
\end{figure*}

The statistical and systematic uncertainties impacting the final extracted cross
sections vary across the unfolded $p_\mu$ and $\cos\theta_\mu$ range (see
Section~IV of the main text). The fractional uncertainties for each 1D bin in
$p_\mu$ and $\cos\theta_\mu$ are shown in Figs.~\subref*{fig:uncert-1d-pmu} and
\subref*{fig:uncert-1d-ctmu}, respectively. The corresponding uncertainties in the
2D space are shown in Fig.~\ref{fig:uncert-2d}.

\section{Additional smearing matrices}

\begin{figure}[h!]
\centering
\subfloat[\label{fig:ac:1d:pmu}]{%
\includegraphics[width=0.48\textwidth]{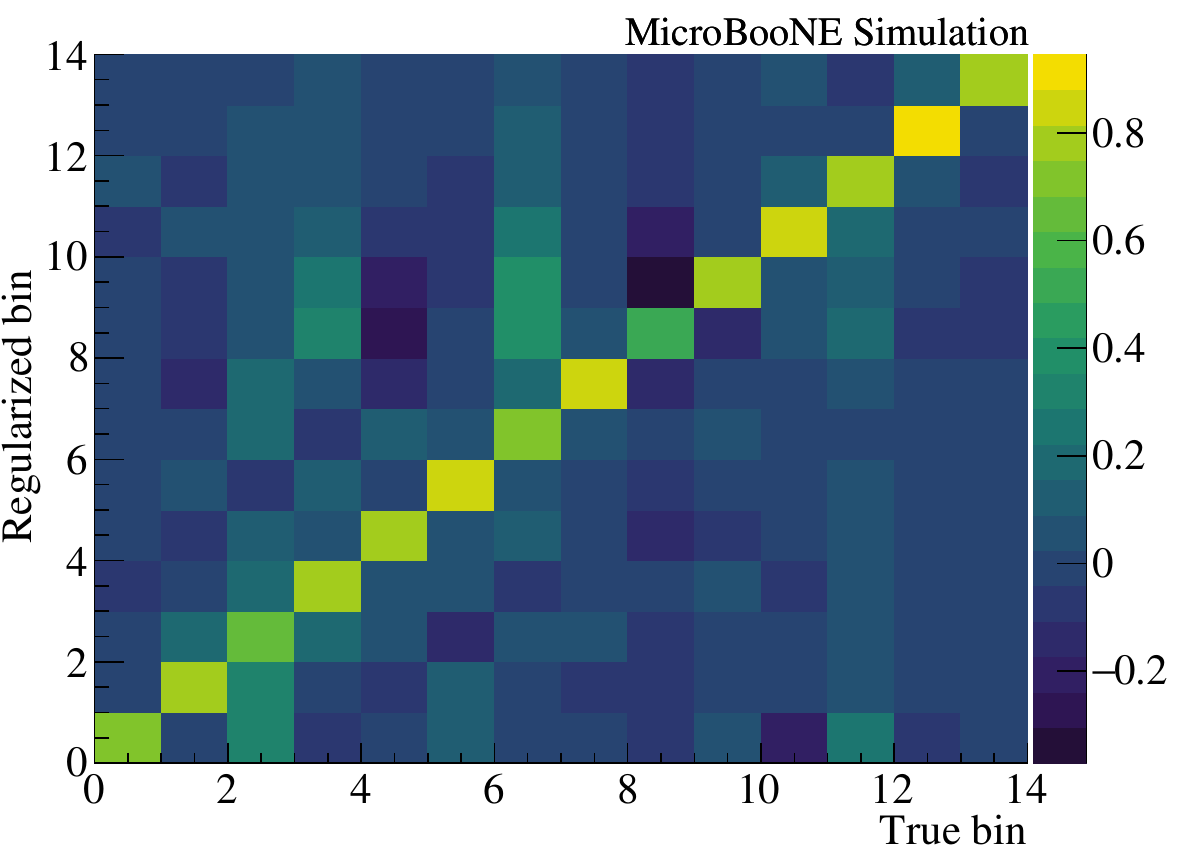}}\\
\subfloat[\label{fig:ac:1d:ctmu}]{%
\includegraphics[width=0.48\textwidth]{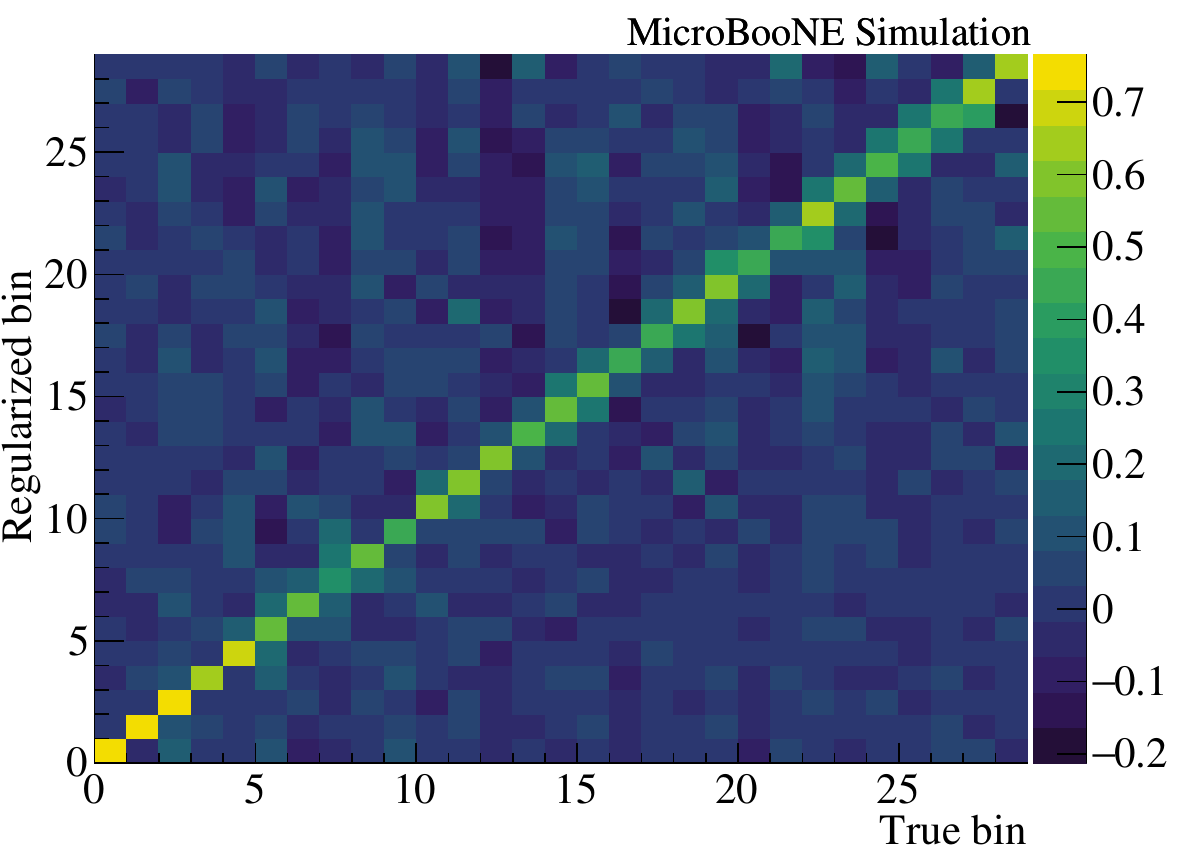}}\\
\subfloat[\label{fig:ac:2d}]{%
\includegraphics[width=0.48\textwidth]{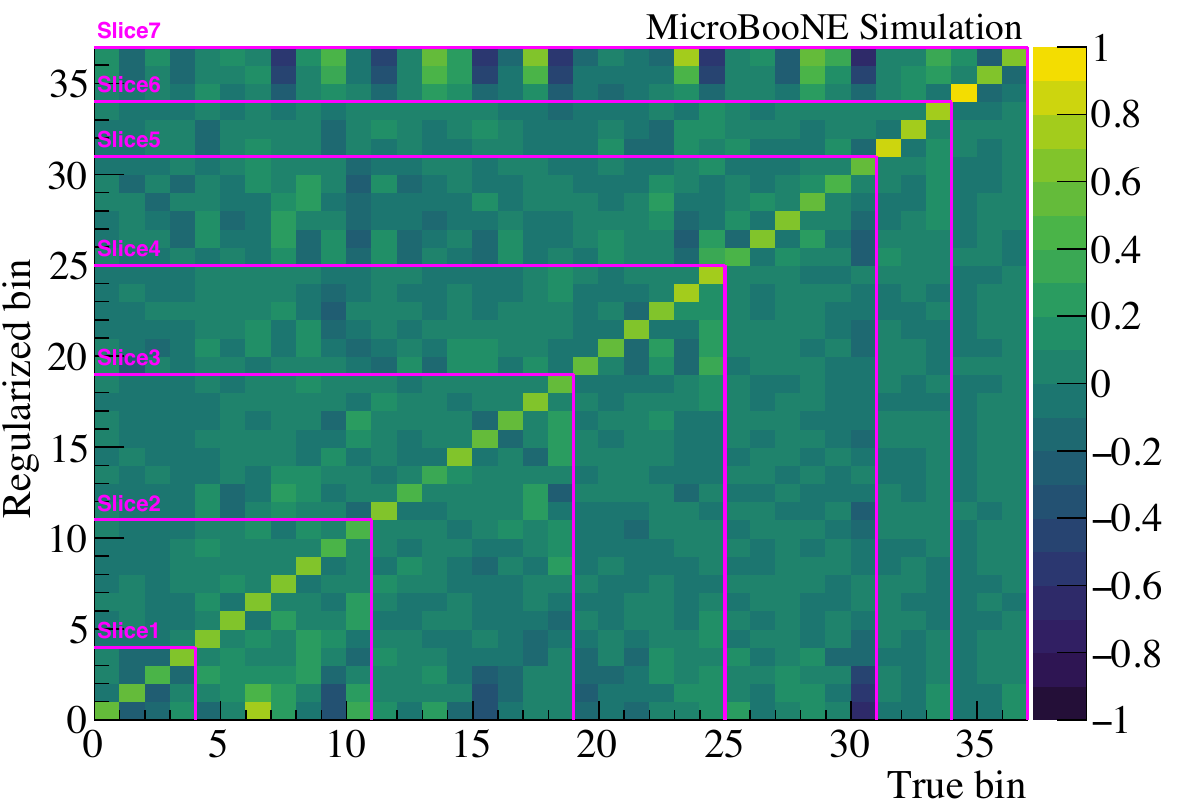}}
\caption{Wiener-SVD additional smearing matrices $A_c$ for each extracted
differential cross section for (a) 1D $p_\mu$,
(b) 1D $\cos\theta_\mu$, and (c) 2D $p_\mu-\cos\theta_\mu$.}
\label{fig:smearing}
\end{figure}

The cross sections extracted using the Wiener-SVD unfolding are in a space of
regularized true observables $\vec{y}$. The smearing matrix $A_c$ encodes the
transformation of cross sections from true observables to this regularized
space. To perform statistical comparisons of generator predictions to the
measured cross sections, the histogram $\vec{x}$ of cross sections in terms of
true observables produced by the generator model must be transformed by the
matrix as $\vec{y}=A_c\cdot\vec{x}$. Here, $\vec{y}$ and $\vec{x}$ represent the
total cross section in each bin, prior to dividing by the bin widths to obtain
the differential cross sections. The smearing matrices $A_c$ are shown in
Fig.~\ref{fig:smearing}.

\section{Generator model comparisons}

We compare the extracted cross sections to several generators and physics models
within GENIE. Here, we briefly describe the main differences in the other
configurations and generators, relative to the baseline GENIE G18-10a model.

We begin with the quasielastic (QE) modeling and initial nucleon treatment,
where further details can be found in Refs.~\cite{Hayato:2021heg,
Golan:2012rfa}. While \texttt{NEUT} and \texttt{NuWro} share the same models for
RES (RS model) and DIS (BY model) interactions as the GENIE CMC tune
\texttt{G18-10a}, they differ primarily in QE modeling and the treatment of the
initial nucleon. Both generators implement the Valencia $2p2h$ model
\cite{Hayato:2021heg} and use the LS formalism for QE interactions, along with
the BBBA05 parameterization~\cite{Bradford:2006yz} for vector form factors. They
also adopt a dipole functional form for the axial form factor based on BBBA07
\cite{Bodek:2007ym}, with \texttt{NEUT} using an axial mass of $M_A^{\text{QE}}
= 1050$\,MeV and \texttt{NuWro} using  $M_A^{\text{QE}} = 1030$\,MeV.

For initial nucleon modeling, \texttt{NEUT} applies the spectral function by
Benhar {\it et al.}~\cite{Benhar:1994hw} in the QE region and the LFG model for other
interactions, while \texttt{NuWro} exclusively uses the LFG model. Additionally,
\texttt{NuWro} incorporates RPA corrections, Pauli blocking, and MEC
contributions~\cite{Golan:2012wx}. The \texttt{GiBUU} model employs a fully
relativistic formalism with parameterized form factors to model neutrino-induced
nucleon and $N$-$\Delta$ transitions. It uses a dipole axial form factor with
$M_A=1000$\,MeV~\cite{Leitner:2006ww}. The impulse approximation is applied to
model the neutrino interaction, followed by the Boltzmann-Uehling-Uhlenbeck
(BUU) transport equation~\cite{Blaettel:1993uz} to describe particle propagation
through the nucleus. The local Fermi gas (LFG) model is used for the nucleon
momentum distribution, incorporating in-medium modifications due to Fermi
motion, Pauli blocking, nuclear binding, and collisional broadening of
resonances~\cite{Leitner:2006ww}. The DIS interactions are modeled by PYTHIA
\cite{Sjostrand:2006za, Mosel:2023zek}, while an empirical MEC model and a
spin-dependent resonance amplitude calculation based on MAID analysis
\cite{Mosel:2019vhx} are also included. We consider two versions of
\texttt{GiBUU}: 2023.3 and 2025. The latter features updated physics libraries,
minor improvements, and bug fixes. For the treatment of FSI, GENIE G18-10a uses
the hA2018 FSI model, an effective single-step approach, with interaction
outcomes determined by probabilistic parameters set by GENIE extracted from
hadron scattering data. The \texttt{NuWro} and \texttt{NEUT} models use a
semi-classical cascade approach for final state interactions (FSI), where
particles propagate through the nucleus in multiple steps, allowing for multiple
interactions while final state hadrons exit the nucleus; this is similar to
GENIE's hN2018 model. Other GENIE tunes using the hA2018 FSI model include
\texttt{G18-10i}, and \texttt{G21-11a}, while GENIE tunes using the hN2018 FSI
model include \texttt{G18-10b}, \texttt{G18-10j}, and \texttt{G21-11b}. Further
details on their differences in FSI approaches between GENIE, \texttt{NuWro},
and \texttt{NEUT} can be found in Ref.~\cite{Dytman:2021ohr}.

To help isolate model differences relative to model implementation differences
across generators, we consider several variations on the GENIE model
configuration. The GENIE CMC tunes \texttt{G18-10i} and \texttt{G18-10j} have
the same model configurations as \texttt{G18-10a} and \texttt{G18-10b},
respectively, but use the $z$-expansion~\cite{Meyer:2016oeg} for modeling the
axial vector form factor in the QE model, in contrast to the dipole form factor
used in \texttt{G18-10a} and \texttt{G18-10b}. The CMC \texttt{G18-10a},
\texttt{G18-10b}, \texttt{G18-10i}, and \texttt{G18-10j} models use the
Nieves-Amaro-Valverde (NAV) QE interaction model~\cite{Nieves:2011pp} and the
Valencia model for $2p2h$ interactions~\cite{Schwehr:2016pvn}. Finally, CMCs
\texttt{G21-11a} and \texttt{G21-11b} use the SuSAv2 model for both QE and
$2p2h$ interactions~\cite{Gonzalez-Rosa:2022ltp}. Results are summarized in
Table~\ref{tab:chi2_combined_full:historical}.

We also consider older versions of the GENIE CMC tunes. The historical
\texttt{G00-00b} CMC represents the default GENIE v2.12 and features the first
(empirical) implementation of the MEC model \cite{Schwehr:2016pvn}. An updated
CMC tune using a similar model set is \texttt{G18-01a}, which introduces
diffractive pion production and hyperon production in the model. CMC tune
\texttt{G18-02a} includes updates to the resonance and coherent pion production
models, transitioning from the RS model to the BS model, which accounts for the
muon mass. The models mentioned above use the LS
formalism~\cite{LlewellynSmith:1971uhs} in the QE interaction region and adopt
the hA2018 FSI model. The \texttt{G00-00b}, \texttt{G18-01a}, and
\texttt{G18-02a} CMCs use the hA2018 FSI model. We observe that the historical
\texttt{G00-00b}, \texttt{G18-01a}, and \texttt{G18-02a} CMCs provide a poor
description of the data. Subsequent model updates, particularly improvements to
the QE and $2p2h$ physics models (i.e., transitioning from LS to NAV or SuSAv2)
yield better agreement. At lower $p_\mu$, generators using a multi-step or
cascade final state interactions (FSI) approach outperform those using a
single-step approach (e.g., GENIE hA). Also, CMCs \texttt{G21-11a} and
\texttt{G21-11b} best describe the data at the lowest $p_\mu$ and at forward
angles, suggesting a preference for the SuSAv2 model in these phase space
regions. Overall, however, the NAV model provides a better description than
SuSAv2, except in the 1D $p_{\mu}$ measurement.

\begin{table*}
\caption{Summary of historical generator model comparisons, indicating the
\(\chi^2\) and $p$~values for each model considered, in comparison to the 2D
extracted cross section and the individual 1D single-differential distributions
in \(p_{\mu}\) and \(\cos\theta_{\mu}\).}
\label{tab:chi2_combined_full:historical}
\centering
\renewcommand{\arraystretch}{1.2}
\setlength{\tabcolsep}{6pt}
\begin{tabular}{|l|c c|c c|c c|}
\hline
\multirow{2}{17ex}{{Model/Tune}} &
\multicolumn{2}{c|}{2D $p_{\mu}/\cos\theta_\mu$} &
\multicolumn{2}{c|}{$p_{\mu}$} &
\multicolumn{2}{c|}{$\cos\theta_{\mu}$} \\  
\cline{2-7} &
$\chi^2/\text{ndf}$ &
$p$~value &
$\chi^2/\text{ndf}$ &
$p$~value &
$\chi^2/\text{ndf}$ &
$p$~value \\  
\hline
\hline
G00-00b & 123.8/37 & 0\% & 12.3/14 & 59\% & 29.1/29 & 46\% \\
\hline
G18-01a & 106.0/37 & 0\% & 17.1/14 & 25\% & 31.0/29 & 36\% \\  
G18-02a & 113.7/37 & 0\% & 19.9/14 & 13\% & 32.8/29 & 29\% \\  
\hline
G18-10b &  63.9/37 & 0\% & 15.0/14 & 38\% & 25.6/29 & 65\% \\ 
\hline
G18-10i &  66.5/37 & 0\% & 18.0/14 & 21\% & 28.2/29 & 51\% \\  
G18-10j &  65.0/37 & 0\% & 15.6/14 & 34\% & 26.1/29 & 62\% \\
\hline
G21-11a &  66.2/37 & 0\% & 11.2/14 & 67\% & 23.8/29 & 74\% \\  
G21-11b &  75.6/37 & 0\% &  9.75/14 & 78\% & 23.9/29 & 73\% \\ 
\hline
AR23-20i &  65.9/37 & 0\% &  16.3/14 & 30\% & 24.2/29 & 72\% \\ 
\hline
\end{tabular}
\end{table*}

\bibliography{references}